\documentclass[conference]{IEEEtran}
\IEEEoverridecommandlockouts
\usepackage{cite}
\usepackage{amsmath,amssymb,amsfonts}
\usepackage{algorithmic}
\usepackage{graphicx}
\usepackage{textcomp}
\usepackage{xcolor}
\usepackage{multirow}
\usepackage{mwe}
\usepackage{mathtools}

\usepackage[ruled,vlined]{algorithm2e}
\usepackage{subcaption}
\def\BibTeX{{\rm B\kern-.05em{\sc i\kern-.025em b}\kern-.08em
    T\kern-.1667em\lower.7ex\hbox{E}\kern-.125emX}}

\newcommand{\sysname}{PNW}
\usepackage{xcolor}
\newcommand{\cut}[1]{{}}

\begin{document}

\title{Predict and Write: Using K-Means Clustering to Extend the Lifetime of NVM Storage\\
}

\author{\IEEEauthorblockN{Saeed Kargar}
\IEEEauthorblockA{\textit{UC Santa Cruz} \\
skargar@ucsc.edu}
\and
\IEEEauthorblockN{Heiner Litz}
\IEEEauthorblockA{\textit{UC Santa Cruz} \\
hlitz@ucsc.edu}
\and
\IEEEauthorblockN{Faisal Nawab}
\IEEEauthorblockA{\textit{UC Santa Cruz} \\
fnawab@ucsc.edu}
}

\maketitle

\begin{abstract}
\cut{
New emerging Non-volatile memory (NVM) technologies are improving the performance and cost-efficiency of computer systems. Nevertheless, their limitations, especially in terms of write endurance, pose new challenges for using them as next-generation memory systems.
}
Non-volatile memory (NVM) technologies suffer from limited write endurance.
To address this challenge, we propose \emph{Predict and Write} ({\sysname}), a K/V-store that uses a clustering-based machine learning approach to extend the lifetime of NVMs.
{\sysname} decreases the number of bit flips for PUT/UPDATE operations by determining the best memory location an updated value should be written to. {\sysname} leverages the indirection level of K/V-stores to freely choose the target memory location for any given write based on its value. {\sysname} organizes NVM addresses in a \emph{dynamic address pool} clustered by the similarity of the data values they refer to.
We show that, by choosing the right target memory location for a given PUT/UPDATE operation, the number of total bit flips and cache lines can be reduced by up to 85\% and 56\% over the state of the art.
\cut{
We design our system to work on hybrid memory systems, where both NVM and DRAM sit on the memory bus. In {\sysname}'s design, the machine learning model and indexing data structure are placed in DRAM and data items are written directly on NVM. 
}

\end{abstract}


\section{Introduction} \label{introduction}

\cut{
In recent years, there has been a growing demand for Storage Class Memories (SCM),  such as Phase-Change Memory (PCM), resistive RAM (Re-RAM) and spin-transfer torque RAM (STT-RAM). Due to their unique characteristics such as non-volatility, high density, high scalability, and byte addressability~\cite{lee2009architecting, burr2008overview, hameed2017efficient, kotra2016re}, they show promise to fundamentally change the existing memory hierarchy. Unlike popular flash memory devices with block erase and program requirements, NVMs support in-place programmability without the need for the erase operation~\cite{cho2009flip,venkataraman2011consistent, xie2011modeling}. 
} 
In recent years, there has been a growing interest in Non-Volatile Memory (NVM)---such as Phase-Change Memory (PCM)---due to their unique characteristics, including non-volatility, high density, high scalability, and byte addressability.
However, these emerging NVMs also pose a number of challenges: They have limited write endurance and asymmetric read/write access properties, requiring special treatment when deployed in large scale computing systems~\cite{lee2009architecting}. While DRAM's write endurance (the number of writes that can be applied to a block of storage media before it becomes unreliable) is on the order of $10^{15}$ writes, NVM technologies, such as PCM, can be written only up to $10^{8}$--$10^{9}$ times~\cite{mittal2015survey}. 
\cut{
Furthermore, asymmetric properties mean that while reads are fast and harmless, writes are slow and harmful (e.g. read latency of PCM is 50 ns versus 1 $\mu$s to write~\cite{lee2009architecting, dong2015minimizing}). 
}



The limited endurance of PCM means that cells can only be written a limited number of times before they ``wear out''. Some recent technologies, such as Intel's Optane DIMM \cite{Optane}, aim to increase the endurance of NVMs significantly. However, unlike other non-volatile technologies such as flash, PCM cells are written on the byte of cache line granularity instead of the page granularity leading to uneven wear-out even on the sub page-level. To ensure failure-atomicity for the data structures stored in NVMs, software schemes, such as logging~\cite{coburn2011nv} and shadowing\cite{ni2019ssp}, are used. This causes extra overheads in terms of write amplification due to writing log entries or creating additional copies of the data~\cite{liu2020lb+}. Even in modern NVM devices, such as Intel's 3DXPoint---where it is claimed that performance is unaffected by the number of modified words in a cache line---it is beneficial to reduce the number of write operations, to improve endurance and retention.

There exists many proposals to increase the write endurance of NVM storage. One promising approach is the Read-Before-Write~(RBW) technique
, in which the content of an old memory block is read before it is overwritten with the new data. This technique replaces each NVM write operation with a more efficient read-modify-write operation. Reading before writing allows comparing the bits of the old and new data, updating only the bits that differ. 
\cut{
This approach can reduce the number of bit flips significantly. In addition to increasing write endurance, reducing the number of bit flips improves performance and energy efficiency, which is especially important for mobile systems with PRAM storage~\cite{cho2009flip}.
Several methods adopt the RBW technique to improve the write endurance of NVMs. For instance, Flip-N-Write (FNW)~\cite{cho2009flip} and Flip-mirror-rotate~\cite{palangappa2015flip} use this technique to flip the new data if it leads to reducing the number of bit flips. Other methods, such as Captopril \cite{jalili2016captopril}, use RBW to mask fixed hot positions to achieve the same goal. 
}
Other proposed methods, such as \cite{liu2020lb+, kannan2018redesigning}, overcome the limitations of NVMs by designing data structures that decrease write amplification.

However, prior methods are either \emph{(1)~application-agnostic} without the ability to leverage the write and data patterns of applications, or \emph{(2)~specialized solutions} that are built for the write and data patterns of specific applications. Particularly, application-agnostic solutions such as FNW~\cite{cho2009flip} and NVM data structures~\cite{liu2020lb+,kannan2018redesigning}, do not leverage the write and data patterns of the application and miss the opportunity to judiciously place writes on memory locations that would minimize bit flips. On the other hand, specialized solutions such as~\cite{jalili2016captopril}, try to minimize the number of bit flips via fixed bit masks that target specific predefined workloads. This renders these solutions limited to predefined applications limiting performance for applications with dynamically changing write patterns.

In this paper, we propose Predict and Write ({\sysname}), an NVM-based K/V store that uses a dynamic approach to minimize bit flips adapting to new applications and dynamic workload changes. {\sysname} decreases both the number of NVM line writes as well as the number of NVM word writes (see section~\ref{Overview and setup}).

We leverage machine learning (ML) to continuously learn a model that reflects the existing write patterns of a given workload. 
The model learns to cluster memory locations in NVM enabling the placement of future writes to locations that minimize the amount of bit flips. Furthermore, by periodically retraining the ML model, it adapts dynamically to a changing workload without the need for user intervention. It is worth noting that unlike the previous methods, which are based on the RBW technique, {\sysname} does not depend on NVM hardware modifications. This is because we do not require using hardware-based read-modify-write operations before write operations as we can avoid writing similar data at a larger granularity (\emph{e.g.} a cache line). However, future work on combining {\sysname} with custom hardware support could further reduce the number bit flips at the bit or byte level.

\begin{center}
\begin{table}
  \begin{center}
    \caption{Comparison of memory technologies \cite{kamath2019storage, van2019persistent}}
    \label{tab:memory_characteristics}
    \begin{tabular}{l|c|c|c}
      \textbf{Category} & \textbf{Read Latency} & \textbf{Write Latency} & \textbf{Write Endurance}\\ 
      \hline
      HDD & 5ms & 5ms & $\geq$ $10^{15}$ \\ 
      DRAM & 50 $\sim$ 60$n$s & 50 $\sim$ 60$n$s & $\geq$ $10^{16}$\\ 
      PCM & 50 $\sim$ 70$n$s & 120 $\sim$ 150$n$s & $10^{8}$ $\sim$ $10^{9}$\\
      ReRAM & 10$n$s & 50$n$s & $10^{11}$\\ 
      SLC Flash & 25$\mu$s & 500$\mu$s & $10^{4}$ $\sim$ $10^{5}$\\ 
      STT-RAM & 10 $\sim$ 35$n$s & 50$n$s & $\geq$ $10^{15}$\\ 
    \end{tabular}
    \label{table:memory_characteristics}
  \end{center}
\end{table}
\end{center}

Our design consists of a ML model, a hash index, a table for storing metadata named the \emph{dynamic address pool} and a data zone to store the actual data or K/V pairs (see section~\ref{Overview and system model}). We also show in the evaluation section that the performance benefits obtained from the ML technique significantly outweigh its overhead in terms of space cost and time. This is the case even when the ML models are running on CPUs without using specialized hardware. Future extensions of our proposal to use methods that process the ML model on specialized hardware such as accelerators and TPUs would further improve the efficiency of our approach~\cite{kraska2018case}.



\cut{
Our experimental evaluation shows that {\sysname} reduces the number of bit flips by up to 85\% when compared with the baseline and state-of-the-art designs on both real-world and synthetic workloads.

Our contribution can be summarized as follows:
\begin{itemize}
  \item To the best of our knowledge, our proposal {\sysname} is the first to utilize machine learning to decrease the number of bit-flips on NVMs.
  \item {\sysname} evenly distributes writes across all the available addresses and memory bits on NVM.
  \item {\sysname} is highly scalable due to handling the meta data on DRAM instead of NVM. It means that we can easily expand the NVM data zone, where the data is stored, without any need to move items around.
\end{itemize}
}

\section{Background} \label{Background}
\subsection{Non-Volatile Memory Technologies}

Emerging Non-Volatile Memory (NVM) technologies, such as Phase-Change Random Access Memory (PCRAM, PRAM, or PCM) and Resistive Memory (ReRAM), provides fast persistent storage, significantly outperforming traditional Disk and Flash technologies. Table~\ref{table:memory_characteristics} shows the performance characteristics of some prevalent memory technologies. While NVM provides similar read latency to DRAM, its write latency is higher than DRAM and thus, minimizing write operations becomes critical for designing software systems on top of NVMs.
%
For this work, we assume a hybrid memory architecture,
where both DRAM and PCM exist on the same main memory level, managed under a single physical address space~\cite{dhiman2009pdram}. Although {\sysname} can support other memory architectures, it is designed to work on hybrid memory systems, in which case NVM acts as a fast persistent memory directly connected to the memory bus.
\cut{
In this architecture, since NVM memory provides low latency and is on the processor’s memory bus, software can access memory directly via loads and stores. 
}

A PCM write operation demands significantly more current and power than a read operation. This property is of great importance in systems like mobile systems, even requiring them to support ``iterative writing'' of data units of smaller sizes than memory words to limit the instantaneous current. For example, in~\cite{kang20060} and~\cite{cho200790}, the write modes of ×2, ×4, and ×8 are supported instead of faster modes like ×16. As another example, in~\cite{hanzawa2007512kb}, the serial writing of even one bit at a time is supported. Integrating NVMs into existing computer systems requires to develop new NVM-friendly data structures \cite{nawab2017dali} that focus on special properties such as reducing write amplification~\cite{kannan2018redesigning}, or being lock-free \cite{nawab2015procrastination, nawab2015zero}. For instance, the techniques proposed in~\cite{zuo2017write, oukid2016fptree, kannan2018redesigning} target the reduction of write amplification, leading to the reduction of bit flips. However, these methods lead to the increase in wear-out cost mostly because of overlooking the reduction of bit flips at the expense of the reduction of write amplification. We show in Section~\ref{Experiments} that this problem can also lead to the increase in the number of written cache lines compared to PNW, which focuses on minimizing bit flips.


\section{Related Work} 
\label{Related Work}
\label{section:relatedwork}

Although recent methods have been able to address write endurance by reducing the number of bit flips through the RBW technique and specialized NVM data structures, they leav out significant opportunities for reducing additional bit flips: (1) application-agnostic methods that do not leverage the write and data patterns including RBW based techniques~\cite{cho2009flip,luo2014enhancing,dgien2014compression} and NVM-based structures that aim to reduce write amplification~\cite{liu2020lb+,kannan2018redesigning} miss the opportunity of using existing patterns in the stored data to minimize bit flips. Specifically, writes are generally updated in place, whereas our proposed technique determines the target memory location based on the written data values (2)~specialized methods that are designed for specific workloads, such as Captopril \cite{jalili2016captopril}, decrease the number of bit flips via fixed bit masks. These methods only work on specific workloads and suffer from significant overheads. In particular, the bit-masks are storage space intensive themselves and, furthermore, as the bit masks are determined once, the approach cannot adapt to changing workloads.

\cut{
{\color{blue} In the rest of this section, we describe some of the RBW and NVM data structures described above. We use some of these solutions in comparisons in the evaluation section. }}


%

In~\cite{dgien2014compression}, the authors propose DCW to find common patterns and then compress data to reduce the number of bit flips in SCM. Like FNW, DCW replaces a write operation with an RBW process. 
MinShift \cite{luo2014enhancing} proposes a method to reduce the total number of updated bits to SCMs. The main idea of this method is that if the hamming distance falls between two specific bounds, the new data is rotated to change the hamming distance. Captopril~\cite{jalili2016captopril} is another method that reduces the total number of bit flips by masking specific hot locations that are flipped more than others. However, as we will see in section \ref{Experiments}, this method suffers from relatively high overhead. More importantly, it is specialized and would only work on a predefined application.



\cut{
Specifically, they compare every write with 4 predefined sequences of bits to decide which bits need to be flipped and which ones need to be written in their original form.}



Finally, there is a group of techniques that find the similarity between items through Locality Sensitive Hashing (LSH) \cite{chum2008near, ghasemazar2020thesaurus}. In this technique, each item is transformed to a hash fingerprint (usually using minhash), and later, LSH is applied to it. Since LSH does not preserve the bit-wise similarity among items, it cannot be efficiently used for bit-wise similarity clustering, which is the main purpose of PNW.


\section{Predict and Write} \label{Predict and Write}

Whenever there is a need for updating memory in-place, the number of bit flips depends on the hamming distance between the old data---currently in the memory location---and the new data, which is going to overwrite the memory location. 
{\sysname} reduces bit flips by avoiding in-place updates and, instead, finding a new memory location for each write that would minimize the hamming distance. By placing the write operation in the right memory location that minimizes the hamming distance between the old and the new data, the number of bit flips can be significantly reduced. While promising for reducing bit flips, this technique introduces several challenges. First, it requires an indirection layer to map a logical value to its current physical location. As the write unit size of NVMs is a byte, storing these mappings on the byte level introduces a significant overhead. Second, the technique requires computing the hamming distance between the new (to-be-written) data and all the available physical data locations. Computing the similarity between all locations is prohibitive.

The first challenge is addressed by leveraging a K/V store that already implements an indirection layer to map keys to values.
\cut{
The overhead that {\sysname} introduces is that it turns every \textsf{UPDATE} operation into a pair of \textsf{DELETE} and \textsf{PUT} operations. In Section~\ref{section:experiments}, we show that this introduces only moderate overhead.}
To address the second challenge (finding the right memory location for a write operation to minimize the hamming distance), we introduce a machine learning approach based on k-means clustering.

\begin{table}
\caption {An example of a PCM with 6 elements} \label{tab:table2}
\vspace*{-1mm}
\begin{center}
\begin{tabular}{|c|c|c|}
\hline
{\textbf{Cluster}} & {\textbf{Index}}  & \textbf{Content} \\ \hline
\parbox[t]{2mm}{\multirow{2}{*}{{1}}}   
        & 0 & '0', '0', '0', '0', '0', '1', '1', '1'   \\ \cline{2-3} 
        & 1 & '0', '0', '0', '0', '1', '0', '1', '1'   \\ \hline
\parbox[t]{2mm}{\multirow{2}{*}{{2}}}   
        & 2 & '0', '0', '1', '0', '1', '1', '0', '0'   \\ \cline{2-3} 
        & 3 & '0', '0', '1', '1', '1', '1', '0', '0'   \\ \hline
\parbox[t]{2mm}{\multirow{2}{*}{{3}}}   
        & 4 & '1', '1', '0', '1', '0', '0', '0', '0'   \\ \cline{2-3} 
        & 5 & '0', '1', '1', '1', '0', '0', '0', '0'   \\ \hline
\end{tabular}
\end{center}
\end{table}

\begin{figure} 
         \centering \includegraphics[width=0.9\columnwidth]{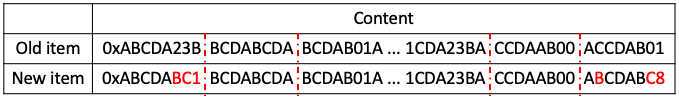}
         \vspace*{-1mm}
         \caption{An example of a memory content that is going to be replaced by a new item with close hamming distance in {\sysname}.}
         \label{fig:cache_line}
 \end{figure}

The intuition behind our clustering approach is that we cluster similar memory locations in terms of the bit patterns of their contents. Using this clustering, we can quickly retrieve a new memory location for a \textsf{PUT} operation such that the hamming distance between the new to-be-written data and the old memory location where it will be written is minimized. We do not need to perform k-means clustering for each \texttt{\textsf{PUT}/\textsf{DELETE}} operation; instead, it is sufficient to perform clustering periodically. We evaluate the training frequency and its effect on reducing bit flips in Section~\ref{section:Training overhead}.

To illustrate our approach, consider a storage system that is using a PCM as its persistent memory with a capacity of six equal sized (8 words) entries, managed by a free-list which we refer to as the \textit{dynamic-address-pool} (Table \ref{tab:table2}). Now, suppose that we have two \textsf{PUT} operations that write the following new data items, d1: ['0', '0', '0', '0', '1', '1', '1', '1'] and d2: ['1', '1', '1', '1', '0', '0', '0', '0']. 

In a regular system, where updates are applied in place, there exists only one option to write the data and hence the reduction of bit flips with techniques such as FNW is limited. {\sysname}, on the other hand, determines the best memory location to write the new data by computing the minimum hamming distance between the new data and existing free memory locations maintained in the dynamic-address-pool. Computing all hamming distances grows in complexity with the number of entries in the dynamic-address-pool and hence becomes intractable. To overcome this problem, {\sysname} groups the entries in the dynamic-address-pool into clusters according to their hamming distance. 

For instance, we can group the elements from the example in Table \ref{tab:table2} into three clusters where indexes 0 and 1 form cluster 1, indexes 2 and 3 form cluster 2, and indexes 4 and 5 form cluster 3. Now, if we receive the same new items d1 and d2, we direct them to clusters that are closest to them, which are clusters 1 and 3, respectively. These items are grouped together because the K-means model assigns data points to a cluster such that the sum of the squared distance between the data points and the cluster’s centroid (arithmetic mean of all the data points that belong to that cluster) is at the minimum. In this example, the centroids for the first, second, and third clusters would be [0.  0.  0.  0.  0.5 0.5 1.  1. ], [0.  0.  1.  0.5 1.  1.  0.  0. ], and [0.5 1.  0.5 1.  0.  0.  0.  0. ], respectively. Because the variations within clusters are minimal, the data points are homogeneous (similar) within the same cluster. In this scenario, wherever we decide to write the items within their corresponding clusters, we will end up writing only 1 bit for each item, without any extra flag bits. This is a simple example of how PNW works. 


\begin{figure} 
\centering
\begin{subfigure}[b]{0.45\textwidth}
   \includegraphics[width=0.9\linewidth]{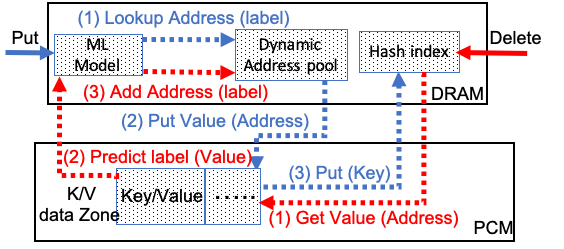}
   \caption{Proposed architecture for small keys}
   \label{fig:put_delete_DRAM} 
\end{subfigure}

\begin{subfigure}[b]{0.45\textwidth}
   \includegraphics[width=0.9\linewidth]{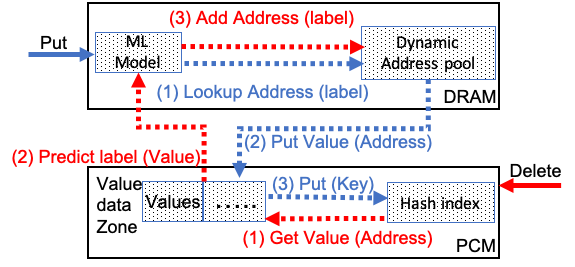}
   \caption{Proposed architecture for large keys}
   \label{fig:put_delete_PCM}
\end{subfigure}

\caption{An example of procedures which serve K/V \textsf{PUT} and \textsf{DELETE} operations for a) small and b) large keys.}
\label{fig:put_delete_operation}
\end{figure}

It is worth noting that {\sysname} reduces the number of writes in two ways: (1) the first way is by writing new items in-place to replace a similar old value in terms of hamming distance. This leads to {\sysname} decreasing NVM word writes (\emph{i.e.}, the number of modified words in a cache line.) (2) In the second way, {\sysname} decreases the number of NVM line writes, respectively cache lines needed to be written per item. For example, suppose that the page size in our system is 4KB as shown in Figure.~\ref{fig:cache_line}. In this scenario, if the items are similar to each other in terms of the hamming distance, fewer number of cache lines are needed to fulfill the request (suppose each part in Figure.~\ref{fig:cache_line} is a cache line). This enables {\sysname} to decrease NVM word writes in addition to NVM line writes.

\cut{
The following summarizes how our method works: we build an ML model stored in DRAM based on the available data items on PCM. When a write operation is received, the ML model predicts the cluster it belongs to. In other words, the new data is directed to a cluster of memory locations with bit patterns that are similar to the value of the write operation (in terms of hamming distance). The write operation is then applied to one of these (similar) memory locations which leads to reducing the number of bit flips. In the next section, we present the details of the algorithms and designs of our proposal.
}


\section{Key-Value Store Design} \label{Design}

In this section, we present the design of our K/V store utilizing the Predict-and-Write technique. We first describe the ML model and then discuss the capabilities supported by our proposed K/V store.

\subsection{Overview and system model} \label{Overview and system model}

Our design consists of a ML model, a hash index, a table for storing available (free) NVM addresses \emph{dynamic address pool}, and the \emph{K/V data zone} to store the K-V pairs. In Figure.~\ref{fig:put_delete_operation}, we show a K/V store on a DRAM-NVM hybrid memory layout using our {\sysname} method. Our data store implementation supports K/V operations including \textsf{GET}, \textsf{PUT}, and \textsf{DELETE}.

\vspace*{0mm}
\begin{figure} 
         \centering \includegraphics[width=0.9\columnwidth, angle =270, scale = 0.8]{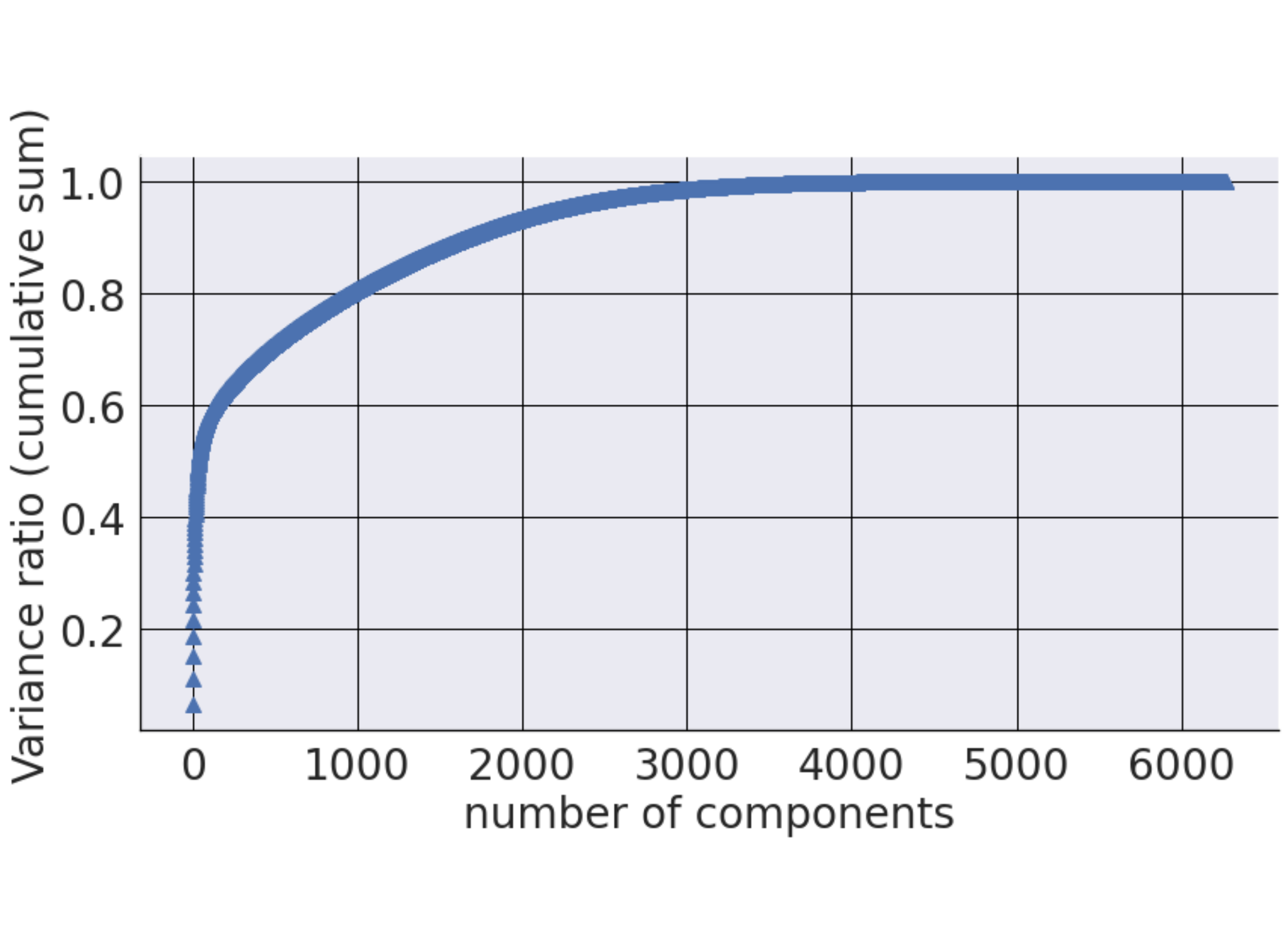}
         \vspace*{-8mm}
         \caption{PCA variance ratio according to the number of principal components.}
         \label{fig:PCA}
\end{figure}

\subsubsection{Machine Learning Model} \label{Machine Learning Model}

Our proposed machine learning method learns the existing data distribution among real-world workloads to decrease the bit flips in write operations. We utilize an unsupervised ML model that is able to cluster data elements into a number of clusters based on their similarity. In particular, we leverage K-means clustering to cluster the available data on PCM. The size of the buckets (the unit of the value size) can vary ranging from a word size to the size of a page or even the size of a document depending on the system.

In our system, each memory location is encoded as a vector of bits, each of which is used as a feature/dimension. The entire data zone can be encoded as a 2D tensor (that is, an array of vectors) of shape (n, m), where the first axis (n) represents the samples (old data) and the second axis (m) represents the features. Because the size of the buckets can be very large (thousands of bits), it can lead to a problem referred to as the ``curse of dimensionality'', which increases the training time and space complexity of the model significantly.

\cut{
To tackle this problem, dimensionality-reduction algorithms are typically used such as principal component analysis (PCA) or t-distributed stochastic neighbor embedding (t-SNE). In this paper, for some data sets with large number of dimensions in their feature space, such as image or video data sets, we use PCA to reduce the dimensions before training the model. It is worth mentioning that after applying PCA to the input, the feature vectors are still the same (binary) with only fewer number of dimensions. For example, in the image and video data sets, we could decrease the number of dimensions of hundreds of thousands to a lower-dimensional space, for example, a couple of hundreds, in such a way that the variance of the data in the low-dimensional representation is maximized (we caught more than \%90 of the variance).
For ease of exposition, we assume that the size of buckets is fixed (multiple sizes can be supported by using different models for each bucket size.)}

\textbf{Addressing the Curse of Dimensionality}
To tackle the curse of dimensionality problem, we use Principal Component Analysis (PCA) on the data sets used in this paper reducing the number of dimensions before training the model. Although PCA is applicable to all data sets, it is especially useful for the ones with a very large number of features. Projecting data to a lower dimensional subspace is very common in different areas such as meteorology, image processing, and genomics analysis, especially before K-means clustering is applied~\cite{zha2002spectral, ng2002spectral, jolliffe2016principal, ding2004k}.
The main basis of PCA-based dimension reduction is to keep only the principle components (features) which explain the most variance in the original data \cite{cangelosi2007component}. Figure.~\ref{fig:PCA} shows the PCA variance ratio according to the number of principal components for MNIST, which is one of the data sets we use in our tests. In this example, we only keep the first 1000 principal components (features) because they are enough to represent more than 80\% of the variance in the data.

\textbf{Determining the Number of Clusters}
Another important decision that needs to be made before training the model is to determine the number of clusters (K). There are a number of ways to determine the optimal value for K \cite{celebi2013comparative}. In this work, we use one of the most common techniques called the ``elbow method'' \cite{joshi2013modified,syakur2018integration,madhulatha2012overview}. The elbow method is expressed as the following Sum of Squared Error (SSE) \cite{syakur2018integration}:

\vspace*{0mm}
\begin{figure} 
         \centering \includegraphics[width=0.9\columnwidth, angle =270, scale = 0.8]{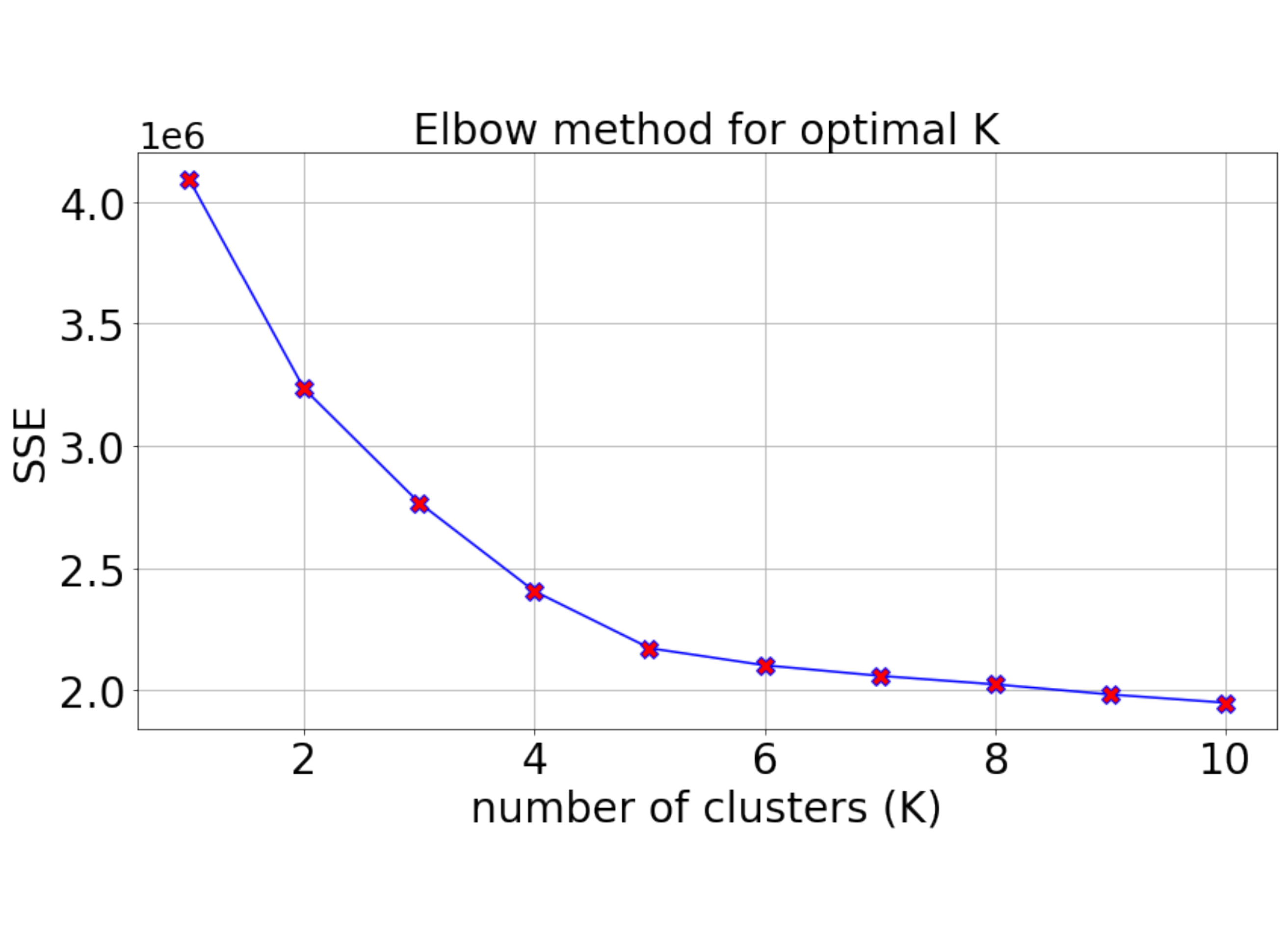}
         \vspace*{-8mm}
         \caption{Sum of Square Error graph to find the optimal K.}
         \label{fig:Elbow}
\end{figure}


\begin{equation}
SSE(X,\Pi) = \sum_{i=1}^{K}\sum_{ \substack{x_{j} \in C_{i} } }
{\lVert x_{j}-m_{i} \rVert_{2}^{2}}
\label{eq:elbow}
\end{equation}

where $\lVert . \rVert_{2}$ denotes the Euclidean (L2) norm, $m_{i}$=$\frac{1}{|C_{i}|}\sum_{\substack{x_{j} \in C_{i}}} x_{j}$ is the centroid of cluster $C_{i}$ where the cardinality is $|C_{i}|$, $\Pi$=\{$C_{1}, C_{2},...,C_{K}$\}, and $X$=\{$x_{1}, ..., x_{i},..., x_{N}$\} (N is the feature vector).

In this method, the value for SSE is calculated as we increase the number of clusters. To determine the optimal number of clusters, we identify a sharp decrease known as the ``elbow'' or ``knee'', which suggests the optimal value for K \cite{syakur2018integration,madhulatha2012overview,vendramin2010relative}. Figure.~\ref{fig:Elbow} shows an example of choosing the optimal K by seeing the significant decrease in the SSE graph, which is in K = 5 (the data set is MNIST).

\cut{
\textcolor{blue} {Another important decision that needs to be made before training the model is the number of clusters (K). Intuitively, by increasing K, we should get better results since the samples within a cluster become more similar to each other in terms of hamming distance, and as a result, PNW evenly distributes writes not only in the address level but also in the bit level (see section~\ref{wear_leveling}). In other words, by increasing K, the number of bit flips decreases. Nevertheless, having more clusters makes the model more complicated. For instance, the model needs more time to be trained when it has more clusters (see section~\ref{Training overhead}). There are a number of methods that can help us to find the optimal value for K, such as the elbow method, the silhouette method, and so on. Although these methods find the optimal K, the final decision is a trade-off that depends on the application that uses PNW.}
}

The ML model is constructed on DRAM as it does not need to be persistent and can be reconstructed after a crash. By constructing the model on DRAM, we take advantage of both DRAM's high write endurance and DRAM's high speed. Another advantage of our proposed method is that this model can be replaced by any customized learning model.









\subsubsection{Dynamic address pool} 
\label{Dynamic address pool}
\begin{figure} 
         \centering \includegraphics[width=0.9\columnwidth]{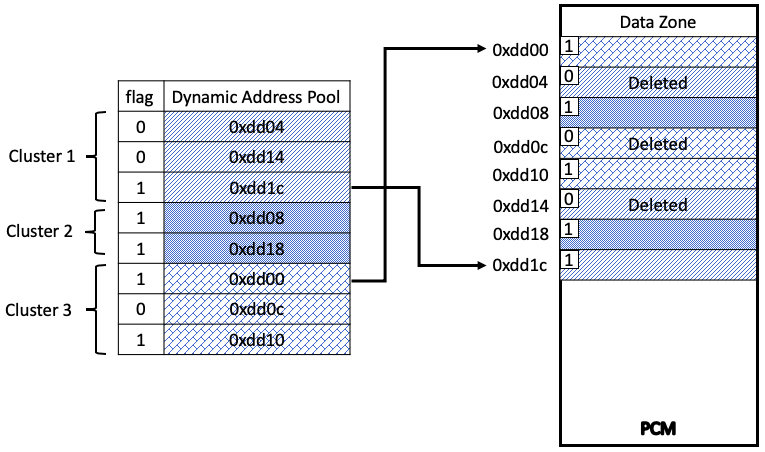}
         \caption{Dynamic Address Pool.}
         \label{fig:DAP}
 \end{figure}

The \emph{dynamic address pool} is a table that contains a number of entries, equal to the number of clusters in the ML model (Figure.~\ref{fig:DAP}). Each entry in the dynamic address pool contains a free-list of the available memory locations that belong to the same cluster, as it is learned by the ML model.


\textbf{Initialization.} The first step of initialization is creating a K-means clustering model based on the number of clusters we want to have, and then training the model on all the available data in the NVM storage called the data zone (Algorithm \ref{alg:init}). The next step is to label data items in each memory location (line 3). Finally, we add the available addresses on the data zone to their corresponding entry in the dynamic address pool (lines 4 and 5). Now, when a \textsf{PUT} request is received by the system, the ML model finds its label, and based on that label, the dynamic address pool returns one address from corresponding cluster. We maintain a flag for each address in the dynamic address pool to indicate whether it is available. We also remove memory addresses out of the dynamic address pool when they are allocated to a K/V pair and reinsert them afterwards to ameliorate the cost of keeping a flag per address in terms of lookup time.

\cut{
by keeping a flag per address, your lookup time increases linearly with the number of taken entries. This has a potentially huge performance impact. This could be ameliorated by removing addresses out of the dynamic address pool when taken and reinserting them afterwards.
}


It is worth noting that the storage overhead of the dynamic address pool is proportional to the number of pointers that are stored per value. As a result, for large values, the size of the table does not grow significantly. For small values, however, the number of addresses that needs to be stored per value can grow substantially. To limit the table size, we set a fixed number of entries in the table, so the size of the table cannot not grow to more than a specific maximum threshold. In this way, the table is used by adding addresses in and removing them from the table until the number of available addresses goes under a minimum threshold, called the load factor, which is described in details in section.~\ref{Other design considerations}.

\begin{algorithm}
{\fontfamily{qcr}\selectfont
        \caption{Initialization}
          // n\_clusters: number of clusters
          
          // D' and A: content and addresses of the data zone
        
          // DAP: Dynamic Address Pool
          
          // N: len(D')
        
        \begin{algorithmic}[1]
        \STATE model = KMeans(n\_clusters) 
        \STATE model.train(D')
        \STATE labels = model.labels\_
        \STATE for (i:=0, i<N, i++)
        \STATE \quad DAP[labels[i]].append(A(i))
        \end{algorithmic}
        \label{alg:init}
}

\end{algorithm}

\subsubsection{Hash index}\label{Hash index}
Indexing is critical in designing K/V stores. Our hash index component maps each key to the memory location that contains its value in the NVM data zone. To build indexes that support K/V operations, there exists a variety of choices ranging from B+-Tree to LSM trees to hashmaps. The operational efficiency of each indexing structure varies from one implementation to another and hence the optimal implementation is application specific. For the existing implementation, we choose hash indexing, however, it can be replaced with any other indexing data structure. The only requirement of the indexing structure is that it can map logical keys to arbitrary physical memory addresses.

We have two choices to store the indexing structure:
\begin{itemize}
    \item If we place the indexing structure into PCM (Figure~\ref{fig:put_delete_PCM}), there is no need to rebuild it during the recovery from a crash. However, it also introduces extra writes to the NVM because of the write amplification problem induced by indexing data structures such as B+Trees and hash indexing. It is a good design choice when the size of the keys are large because in that case the wear-out cost of the hash index is negligible. However, for small keys, it represents a problem which we mitigate by leveraging data structures such as NVM-friendly hashing indexes~\cite{zuo2017write}.
    \item Another design choice is to build the indexing structure on DRAM (Figure~\ref{fig:put_delete_DRAM}). This architecture is particularly beneficial when the size of the keys are small. In this case, we do not pay any cost for the extra bit flipping that is caused by the write amplification of the indexing structures. Nonetheless, we need to build the whole data structure from scratch during recovery after a crash. 
\end{itemize}

\cut{
{\color{blue}
Depending on the placement of the hash index, either that it is sufficient to store values only in the data zone (if the hash index is in NVM) or that is is required to store both keys and values (if the hash index is in DRAM and needs to be recovered after a crash.)
}}

In the evaluation, we build and persist a write-friendly hash index in PCM as introduced in \cite{zuo2017write}. We perform the tests based on this design to explore the worst case scenario of putting the hash index on PCM in terms of extra bit flips introduced by write amplification. Also, for every entry in the hash index, there is a flag bit that shows whether the corresponding key is available or not. In particular, whenever we receive a delete request, we can reset its corresponding bit in the hash index to reflect that the corresponding index does not exist anymore instead of deleting it. We can do the same procedure for deleting a K/V pair from the data zone.

\subsection{Supported K/V Operations}

\subsubsection{\textsf{PUT} Operation}

\textsf{PUT} and \textsf{UPDATE} operations are executed as follows. As shown in Figure~\ref{fig:put_delete_operation}, when our system receives a write request such as \textsf{PUT}, the model is queried to determine the cluster that is closest to the value-to-be-written in terms of their hamming distance. Then, a memory address is returned from that cluster by using the \emph{dynamic address pool}. Then, the K/V pair is written into the returned address, which is in the K/V zone on NVM. Finally, the newly-added index entry is added to the hash index (step 3).

\begin{algorithm}[bp]
{\fontfamily{qcr}\selectfont
        \caption{Write operation}
          // D' and D: old and new (key,value)
        
          // DAP: Dynamic Address Pool
          
          Write (D: (key,value))\{
        
        \begin{algorithmic}[1]
        \STATE E = model.predict(D); //predict the entry
        \STATE A = DAP.get(E);//get the address
        \STATE D'= Read(A); //old (key,value)
        \STATE DAP.remove(A) //remove the address from DAP
        \STATE for each bit in \{D\} and \{D'\}
        \STATE \quad if they differ, update memory bit
        \STATE HI.put(D, A) //update the hash index\}
        \end{algorithmic}
        \label{alg:write}
}
\end{algorithm}

\begin{algorithm}
{\fontfamily{qcr}\selectfont
        \caption{\textsf{DELETE} operation}
          // D': old key
        
          // DAP: Dynamic Address Pool
          
          // HI: Hash Index
          
         Delete (D': key)\{
         
        \begin{algorithmic}[1]
        \STATE A = HI.get(D'); //get the address 
        \STATE Reset-Flag-Bit(A);//delete 
        \STATE E = model.predict(Read(A)); //predict the entry
        \STATE DAP.update(A:address, E:entry);//add the address back to DAP\}
        \end{algorithmic}
        \label{alg:delete}
}
\end{algorithm}

Algorithm~\ref{alg:write} illustrates the pseudo-code of the write operation under the {\sysname} scheme (Figure~\ref{fig:put_delete_operation}). The first step of {\sysname} is to find its label, which is equal to its corresponding entry in dynamic address pool, using the ML model (line 1). Next, we select one of the available addresses from the corresponding entry in the dynamic address pool, and write the data to the address (lines 2 and 3). Next, we need to remove the selected address (A) from the cluster's free-list in the dynamic address pool (line 4). Finally, only the bits (in the buffer {D}) that are different than the data in PCM ({D’}) are actually updated (lines 5 and 6). We also need to update the hash index at the end to enable finding the value for future lookups (line 7).

\subsubsection{\textsf{DELETE} operation}

Algorithm~\ref{alg:delete} illustrates {\sysname}'s delete operation (also see Figure~\ref{fig:put_delete_operation}). The delete procedure is accomplished by the following steps. In step 1, to find the item in the K/V data zone, the delete request is directed to the hash index, and then the associated entry is deleted from the K/V data zone by resetting the associated flag bit (lines 1 and 2). In this step, the delete operation is completed; however, to make the system more efficient, we recycle the recently freed address back to the dynamic address pool by finding the label of the deleted data (line 3), and then adding the freed address to the corresponding entry in the dynamic address pool (line 4). In this way, the address can be used again in the future, and the model is re-trained less frequently.

\subsubsection{\textsf{UPDATE} Operations}

An update operation can be implemented in two different ways: 

\begin{itemize}
    \item If we care about the write endurance more than latency, the update operation consists of the delete operation plus the \textsf{PUT} operation in order to prevent bit flipping as much as possible. It means that the item that has to be updated is first deleted from NVM (delete operation), and then its new place is found by in a dynamic address pool (\textsf{PUT} operation) using the model. It is worth noting that we can do the \textsf{DELETE}-\textsf{PUT} process asynchronously to mitigate the latency problem. In other words, the system can retain synchronous updates to K/V items and the hash index in NVM, and for the dynamic address pool in DRAM, it can be asynchronously updated through the model in the background to hide the extra latency.
    \item On the other hand, if the application cares about latency more than the other factors, especially wear-leveling, the request just needs to go through the \emph{hash index} to find its place in the K/V data zone and then update the item in place without any further changes since it does not affect the dynamic address pool. In this way, we sacrifice wear-leveling to achieve lower latency. 
\end{itemize}

In our system and evaluations, we follow the first approach as our main goal is to increase write endurance. However, it turns out---as we present in experimental evaluations---that minimizing bit flips is also good for performance alleviating the trade-off between write endurance and latency.

\subsubsection{\textsf{GET} Operation}

Read operations in our system are straight-forward as they do not lead to changing any data structures. Specifically, a get request goes through the hash index to find its corresponding value from the K/V data zone, and then the read value is returned.

\subsection{Additional design considerations}\label{Other design considerations}

It is possible that all the available addresses of a cluster (called cluster C) are utilized. In this case, if the model sends a request that requires a new address from cluster C, the dynamic address pool will not be able to serve this request because there are no more addresses available in that cluster. To avoid this problem, we define a load factor for the K/V data zone on the NVM. Setting the load factor to $x$ percent, means that when x percent of the available addresses in the K/V data zone are used, the K/V data zone needs to be extended. To add new memory addresses to the data zone, we need to train a new model. It is worth noting that, unlike traditional methods, we do not need to move or change anything in the hash table on NVM because they still have valid information. The only things that need to be changed are the model and dynamic address pool, which are both located on DRAM. So, our method to expand the size of a cluster does not impose any extra writes to the NVM.

The main reason behind defining the load factor is to prevent latency spikes or stalls in the system. The load factor is similar in principle to the load factors that are used in hashing schemes as a way to monitor the space utilization of the system to prevent hash collisions. In other words, the load factor is going to warn us that the system will need to be retrained in the near future. So, before this happens, we can re-train a new model, by adding some new memory locations to the K/V data zone, in the background while the system is running. Then, we can switch to the new model and table before the previous model gets stuck. In this case, we can hide the re-training latency and the system works without disruptions due to retraining. We have done some tests in the next section to figure out the best time to start training a new model before the old one is full to keep the system working smoothly.
{\sysname} supports any size of key/values from 32-bit word size to the page sizes of 4KB to the size of a document.
Thereby, the way in which data elements are provided to the models depends on the K/V pair size. For instance, small (e.g. 64 bit) data elements can be directly passed to the model, while for large data element (e.g. 4KB) we first apply dimensionality reduction using PCA before passing the data to the model.



\cut{
The following summarizes our design choices:
\begin{itemize}
    \item To achieve a balance between system performance and memory utilization, we keep the fast indexing hash table on slow PCM and the slow ML model on fast DRAM. This design choice allows us to take advantage of the high speed read/write operations of DRAM while making the in-PCM data simple and efficient.
    \item By building the ML model on DRAM, we save PCM from extra write operations during the training phase, which requires a lot of read/write activities. In other words, the training phase of the model happens in DRAM, and then we can save the whole model on either PCM or DRAM.
    \item {\sysname} is designed in a way where the ML model does not have to be persistent and can be recovered after a crash.
\end{itemize}
}

\section{Evaluation} \label{Experiments}
\label{section:experiments}
\subsection{Methodology}\label{Overview and setup}
In this section, we evaluate our proposed method using different metrics focusing on the reduction in writes and bit flips. We leverage a collection of real and synthetic data sets. Since only insert and delete requests cause mutating the state of the NVM, we insert n items into the K/V store followed by deleting 0.5n items (except for section~\ref{Training overhead}). Also, we do not make any assumption about the access pattern within or across clusters. So, we simply apply the K-means clustering (from the scikit-learn library) based on the available memory locations on PCM. We compare PNW with both RBW solutions and K/V stores. For the former, we compare with the writes on the storage component of PNW, which is the data zone.

We compare our results against other methods described in Section \ref{section:relatedwork}, such as FNW \cite{cho2009flip}, DCW \cite{yang2007low}, Captopril \cite{jalili2016captopril} and MinShift \cite{luo2014enhancing}. For synthetic data sets, our sample K/V store system has at least 10M buckets. When there are 10M buckets, for instance, we first warm-up K/V stores with 10M key/values. This means that we store some items as ``old data'' before starting our tests. The data type and distributes of these items differ depending on the test. ``old data'' is used for the initial training of the ML model.

To compare {\sysname}'s results with other methods, we tune their parameters in such a way that they achieve their best performance. For example, we allow MinShift to shift n times, where n is the size of the item instead of the size of the word, which means it always results in its best performance in terms of the number of bit flips \cite{luo2014enhancing}. With respect to Captopril, we also considered its best case, which happens when the blocks are partitioned into n = 16 segments \cite{jalili2016captopril}.

\begin{figure*}[!ht] %
\centering
\begin{subfigure}{.33\textwidth}
\includegraphics[width=\columnwidth, angle =0, scale = 0.9]{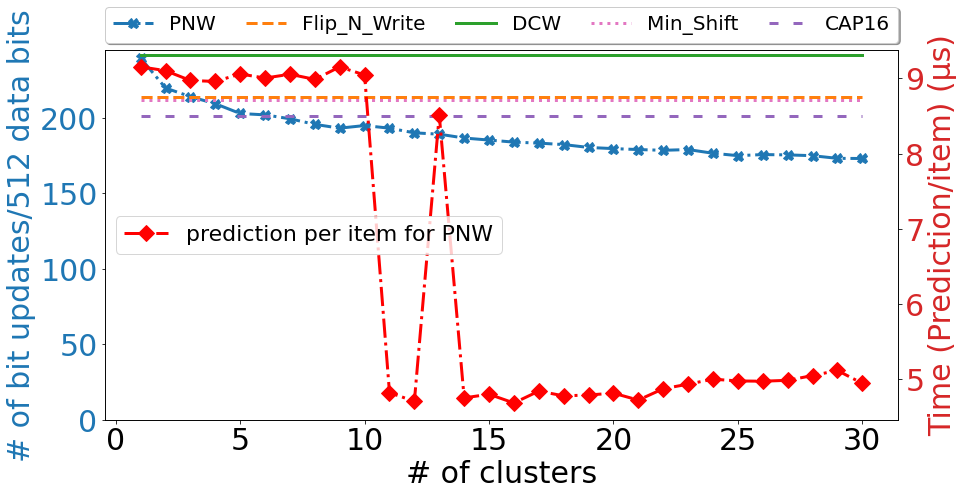}%
\caption{Amazon Access Samples}%
\label{fig:fig_1_Amazon}%
\end{subfigure}\hfill%
\begin{subfigure}{.33\textwidth}
\includegraphics[width=\columnwidth, angle =0, scale = 0.9]{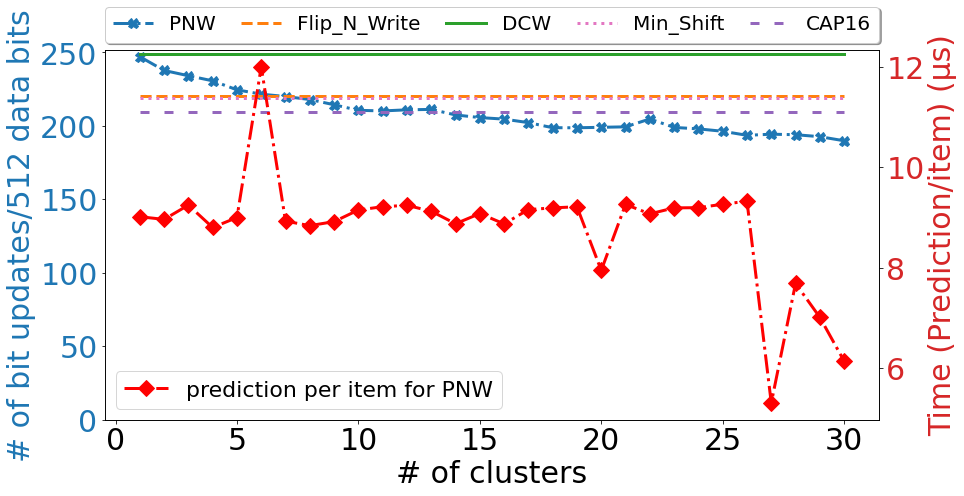}%
\caption{3D Road Network}%
\label{fig:fig_1_3D}%
\end{subfigure}\hfill%
\begin{subfigure}{.33\textwidth}
\includegraphics[width=\columnwidth, angle =0, scale = 0.9]{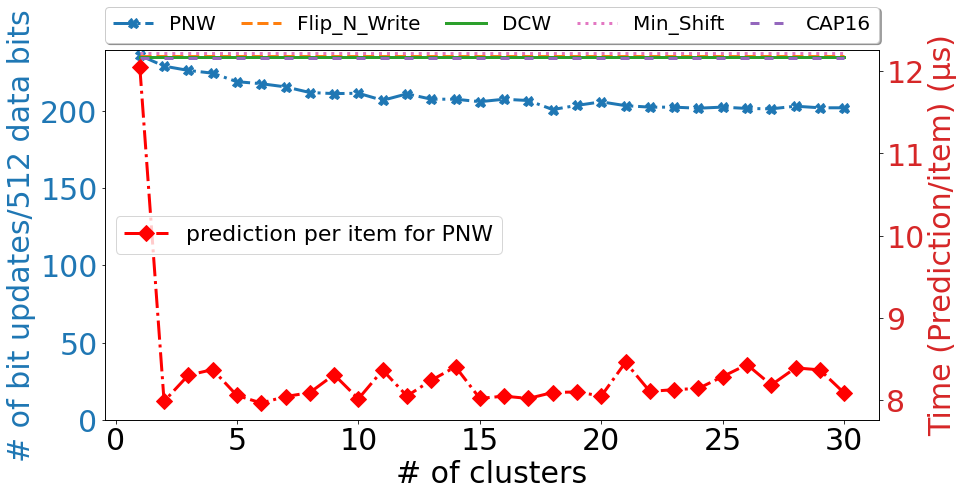}%
\caption{Sherbrooke}%
\label{fig:fig_1_Sherbrooke}%
\end{subfigure}\\
\begin{subfigure}{.33\textwidth}
\includegraphics[width=\columnwidth, angle =0, scale = 0.9]{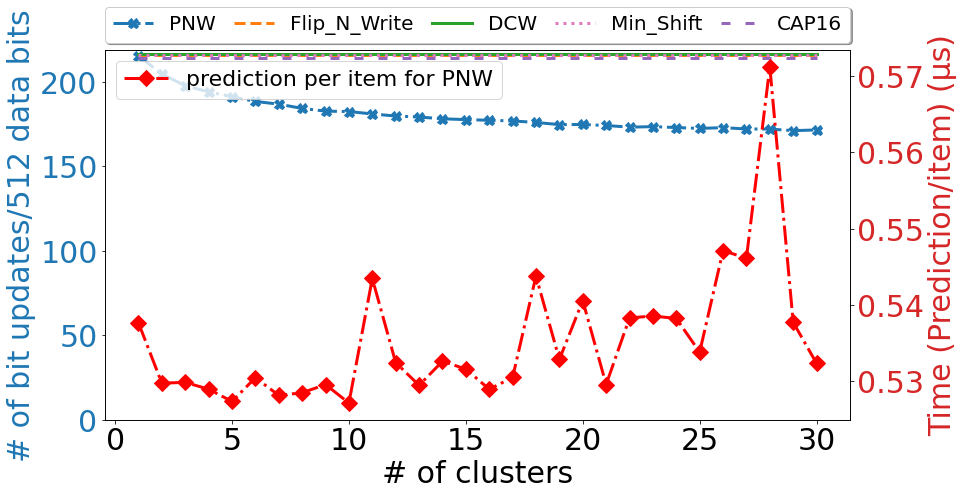}%
\caption{seq 2 traffic surveillance}%
\label{fig:fig_1_traffic}%
\end{subfigure}\hfill%
\begin{subfigure}{.33\textwidth}
\includegraphics[width=\columnwidth, angle =0, scale = 0.9]{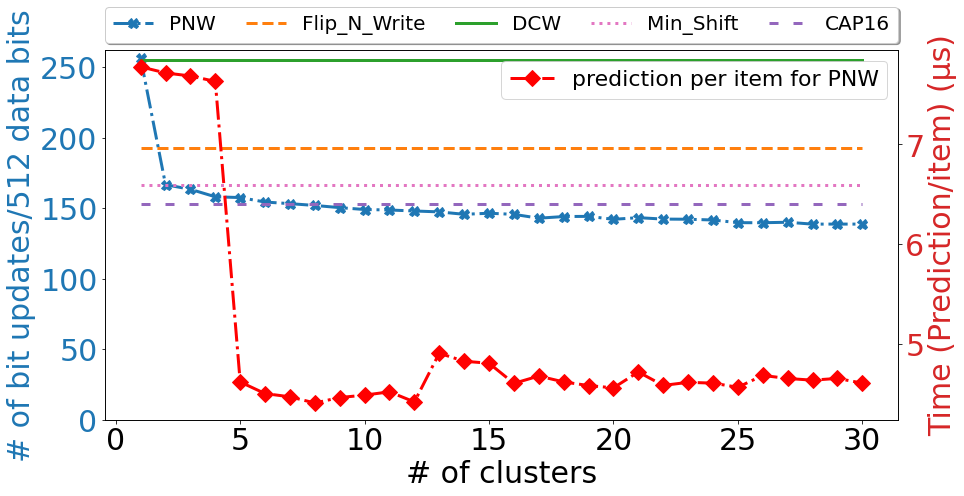}%
\caption{normal data distribution}%
\label{fig:fig_1_normal}%
\end{subfigure}\hfill%
\begin{subfigure}{.33\textwidth}
\includegraphics[width=\columnwidth, angle =0, scale = 0.9]{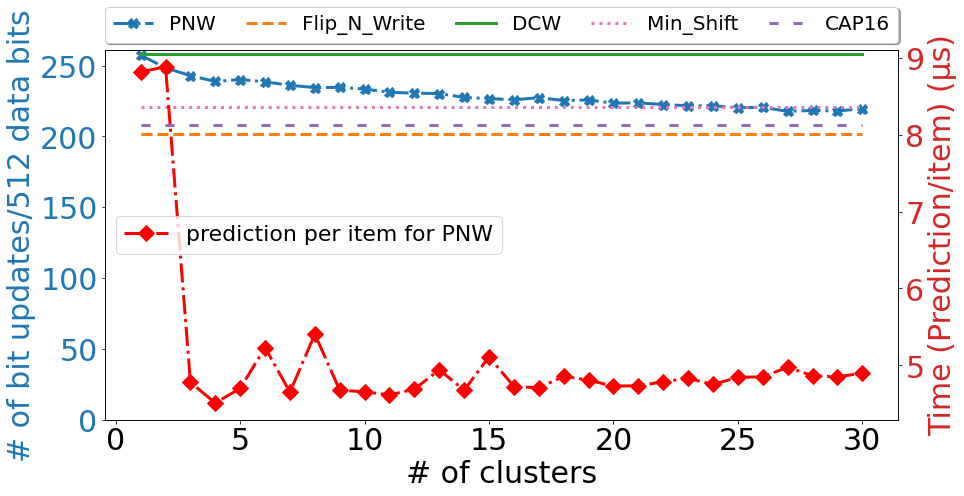}%
\caption{uniform data distribution}%
\label{fig:fig_1_uniform}%
\end{subfigure}\\
\caption{The average number of actual bit updates per writing 512 bits as well as the latency of prediction per item in {\sysname} for the real-world textual and numerical data sets (a-b), multimedia data sets (c-d), and synthetic data sets (e-f).}
\label{fig:fig_realistic}
\end{figure*}

Unless we mention otherwise, we execute the K/V operations with randomly selected key/values from the same generator. As real NVM DIMMs are not available for us yet, we emulate NVM using DRAM similar to prior works \cite{ou2016high, volos2011mnemosyne, huang2014nvram,xia2017hikv}. 
We assume an access latency of the latest 3D-XPoint of 600ns~\cite{izraelevitz2019basic,niclosing}.

The experiments are executed on an Intel Core i7 processor running at 2.2 GHz with 2 cores (4 logical cores), each of which has 256KB L2 Cache and 4MB L3 Cache using 8 GB of RAM, running macOS Catalina (version 10.15.4). The reason that we run the tests on a local computer without any GPU support is to get a sense of how our methods work on an ordinary system without any unique capabilities. We test our proposed method using various data sets, which can be categorized as real-world textual and numerical data, real-world multimedia data including image and video data sets, and finally, hard-to-cluster synthetic data sets. In the following subsections, we show the results of the tests on these data sets and analyze them.

\subsection{Real-world textual and numerical data sets}


The first data set is called Amazon Access Samples Data Set \cite{Amazon_Data_Set, Dua:2019}, containing 30K log entries. Although this data set has 20K attributes, in this test, only less than 10\% of them are used for each sample. For this test, we first have set aside 5K buckets as the ``old data'' on the NVM memory and then warmed up the system by writing 5K items from the data set into our buckets. Then, we replaced this ``old data'' with new incoming data from the same data set (the remaining 25K items). Figure.~\ref{fig:fig_1_Amazon} illustrates that when there are one or two clusters, the number of written bits in our method is more than FNW. Nevertheless, when the number of clusters is more than 2, we start to get better results until we reach between 15\%(compared to CAP16) to 70\%(compared to the conventional method) improvements compared to the other methods when the number of clusters is 30. 

The next real-world data set, i.e., 3D Road Network Data Set \cite{kaul2013building, guo2012ecomark}, contains information of road networks in North Jutland, Denmark (covering a region of 185 x 135 $km^2$). We used the same setup as above for this data set containing 434874 entries. In this test, we chose 100K buckets as ``old memory'' and warmed up the system by 100K entries from the 3D Road Network Data Set. The results are shown in Figure.~\ref{fig:fig_1_3D}. When the number of clusters is big enough (here k = 14), {\sysname} starts to outperform all the other methods in terms of the number of bit flips until it gets its highest performance when k=30 (between 10\% to 63\% improvements compared to the other methods).  

Finally, the last real-world data set is one of the collections of a database called DocWord, which consists of five text collections in the form of ``bags-of-words''. This collection, which is called PubMed abstracts \cite{Dua:2019}, consists of 730 million words in total. For doing the tests, we first created 100M buckets as the ``old data'' storing data from the PubMed data set. Then, we wrote the new incoming data items from the same data set on the previous data items stored on the buckets and kept track of the number of the updated bits per 512 bits. 

\subsection{Real-world multimedia data sets} \label{Real-world multimedia data sets}

In the first set of tests, we have used some video data sets to see what happens if a system, for instance, a CCTV recorder, uses an NVM media as its persistence memory. We have used two video data sets: 1) The Sherbrooke video data set \cite{jodoin2014urban}, representing a two-minute-long video (with resolution 800x600). 2) A Traffic Surveillance video \cite{bahnsen2018rain}, collected from seven intersections in the Danish cities of Aalborg and Viborg, containing 21 five-minute sequences of two cameras including RGB and thermal data. The resolution of both cameras is 640x480 pixels, and the frame rate is fixed at 20 fps. In this test, we just used one sequence of RGB camera called ``day sequence 2''. 

\cut{
\textcolor{blue}
{
Furthermore, since the number of dimensions in these data sets' feature space is very large, we used PCA to reduce the dimensionality before feeding the samples to the model. By doing this, we caught more than 90 percent of the variance existing in these data sets by reducing their dimensionality from hundreds of thousands to a couple of hundreds.}
}

For the first data set (Sherbrooke), we stored the first 30 seconds of this video as the old data, and for the second one (Seq2), we stored the first one minute of the video as the old data and used the remaining of the video as the new data. The results are shown in Figure.~\ref{fig:fig_1_Sherbrooke} and \ref{fig:fig_1_traffic}, respectively. These figures show that our method outperforms the other ones in both data sets. For the first data set (Sherbrooke), {\sysname} improves the other methods between 14\% to 60\% and for the second one (Seq2), we outperforms the other ones between 21\% to 67\%.

The next data set is one of the most widely used data sets for machine learning research, and especially for computer vision algorithms, i.e., CIFAR-10 data set \cite{krizhevsky2009learning}. This data set is a subset of the 80 million tiny images data set and consists of 60,000 32x32 color images, grouped into ten different classes. Similar to the previous experiments, we first set aside 10K of these images as the old data to fill out the 10K buckets we created as our NVM system. Then, the new incoming data items are written in place of the old ones one by one. 

\subsection{Hard-to-cluster synthetic data sets}
\label{sub:synthetic}

In this section, we are going to observe the behavior of {\sysname} on some synthetic data sets that do not follow any specific data distribution. The reason of doing these tests is to discover the limitations of our ML-based method and analyze them to give the readers a clearer view of the possible applications of {\sysname}. To perform these tests, we start with a synthetic data set that shows a clear pattern and then test two more data distributions that are completely different in terms of their data pattern.

For the synthetic data sets, we used 32-bit keys and values. We also generated two types of integer data (normal and uniformly distributed), ranging from 0 to $2^{32}$. For random integers, we generated them via a pseudo-random number generator. For the normal data set, we generated a synthetic data set of 100M unique values sampled from a normal distribution with $\mu$ = $2^{31}$ and $\sigma$= $2^{28}$ to test our method. In all synthetic data set tests, the confidence interval was less than $10^{3}$ for 95\% confidence level.

First, we show the results of the first synthetic data set, following a regular pattern. Figure.~\ref{fig:fig_1_normal} shows the results for different number of clusters ranging from k=1 to k=30 for normal distribution. We have compared the performance of {\sysname} to the other ones in terms of the number of bits updated/written per 512 bits. 
\cut{
Before starting the test, we warmed up the system by generating 10M data items (the old data) from the normal data distribution. Then, we start writing the ``new data'' in place of the ``old ones'' and observing the behavior of the system in terms of the number of bits being updated using different methods. A lower number of bit updates is better as it means less energy consumption, less wear-leveling, less write latency, which in turn means more lifetime, and more write bandwidth. 
}
In this figure, we observe that when we pick k=1, the result for {\sysname} is not different from DCW since both do the same thing if there is no clustering. Our approach enhances the results of DCW and FNW more than 40\% and 25\%, respectively, when the number of clusters is more than 10. It also outperforms MinShift and Captopril more than 15\% and 10\%, respectively. Also, the delay is almost 5$\mu$s to 6$\mu$s most of the time.


In the second experiment, we did the same, but for a different data distribution, i.e. uniform random distribution, to learn more about the behavior of our method. Data sets like this one are highly random, and as a result, difficult to learn using an ML model. The results are depicted in Figure.~\ref{fig:fig_1_uniform} showing that although our method has succeeded in improving the results for DCW, MinShift, and the conventional method by almost 15\%, 5\%, and 60\%, respectively, it lags behind FNW and CAP16 for this data set as expected for the random data set.


In some of the previous results, there are anomalies where the number of bit flips suddenly jumps while increasing the number of clusters. Such anomalies are due to the unpredictability of ML-based methods. However, we expect that such anomalies would be normalized during extended operation.



\begin{figure} 
         \centering \includegraphics[width=0.9\columnwidth, angle =270, scale = 0.8]{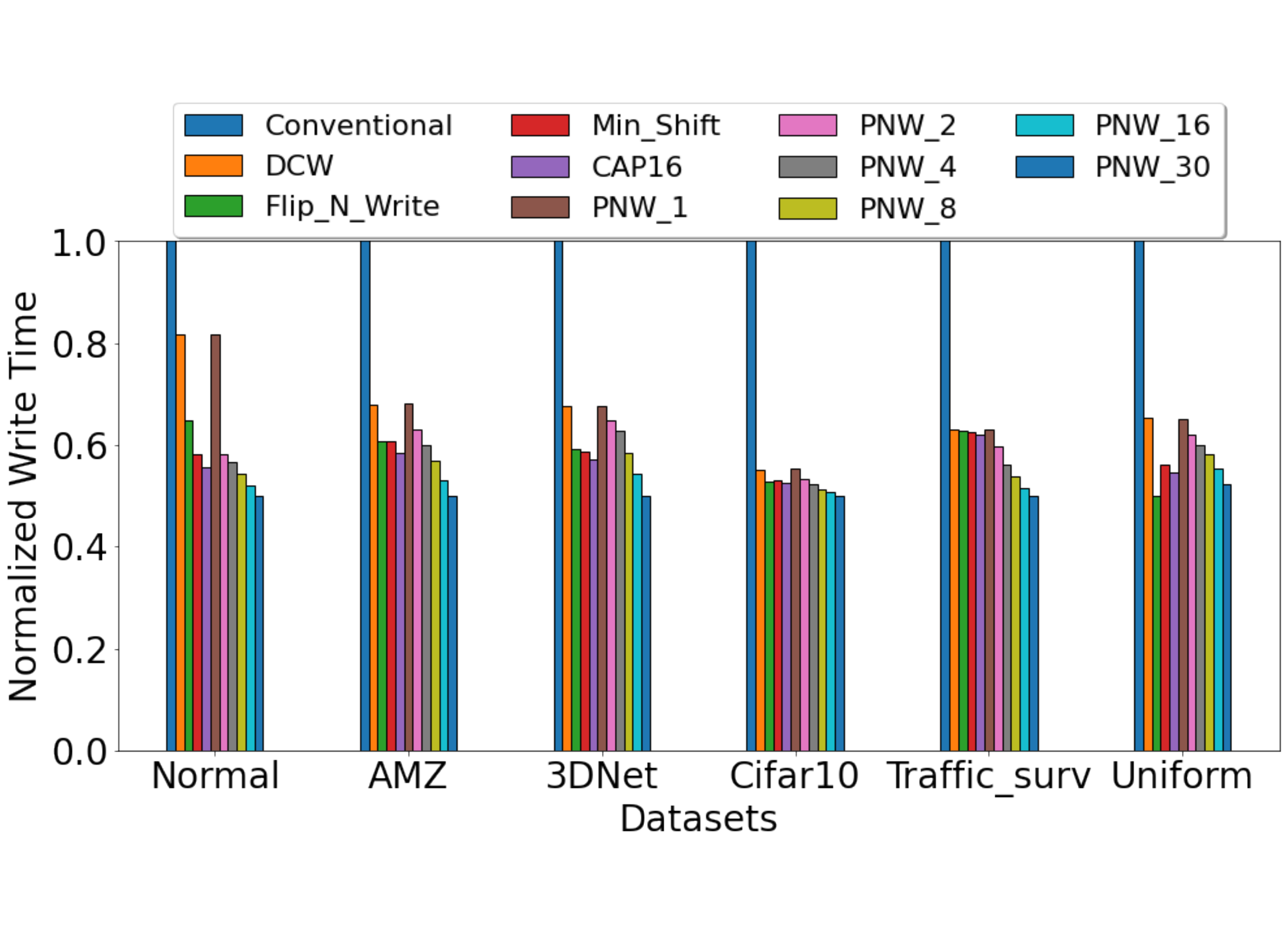}
         \vspace*{-8mm}
         \caption{End-to-end write latency comparison for various data sets.}
         \label{fig:fig_9}
\end{figure}

\subsection{End-to-end write latency} \label{End-to-end write latency}

In the following, we are going to measure the write latency, which includes the time spent on 1) predicting a cluster number, 2) finding an empty bucket within the dynamic address pool, and 3) writing the key/value on NVM. We do this test to measure the overhead of our method.

In Figure.~\ref{fig:fig_9}, we show the write latency comparisons for various data sets. In this test, we use the normal and uniform data distributions, Amazon Access Samples, 3D Road Network, CIFAR, and the \emph{day sequence 2} traffic surveillance video. 
\cut{For this test, we first separate 20 percent of each data set as the old data. Then, we write the remaining 80 percent of each data set on the old data.} For our method, we had to train the model based on the old data, filling out the dynamic address pool, and then writing the new data. The write latency is calculated based on the number of cache lines that are written per item. In this test, we observe that each method that updates fewer bits has a higher chance of having a lower write latency because it has to update fewer cache lines than the others.

Figure.~\ref{fig:fig_9} shows the normalized time of the write operation required by different methods. As illustrated in this figure, our proposed method, when the number of clusters is enough, can outperform the others even though it has to perform two additional steps. The reason is that our method performs fewer write operations than the other ones, and it makes up the time it spends on the extra steps. However, for the uniform data distribution, we could not do the same since {\sysname} is not able to find a clear pattern among the data items to make up the extra steps. 
\cut{
As a result, when there is a pattern, we outperform the other methods when the number of clusters is more than 8. Furthermore, by increasing the number of clusters, we get better results.
}

\begin{figure} 
         \centering \includegraphics[width=0.9\columnwidth, angle =270, scale = 0.8]{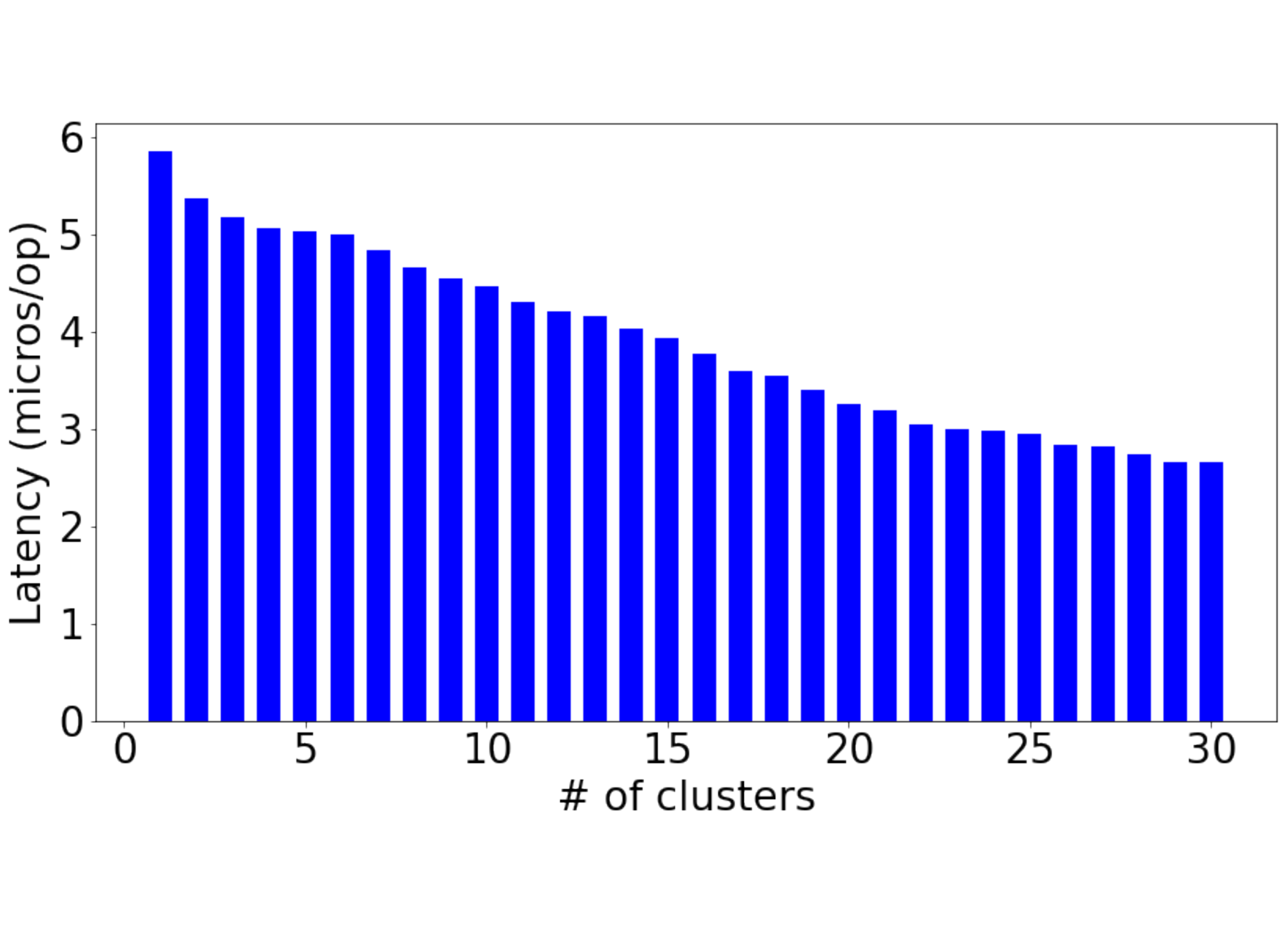}
         \vspace*{-8mm}
         \caption{The impact of choosing the number of clusters (K) on the average write latency for the PubMed abstracts data set.}
         \label{fig:write_latency}
\end{figure}

Figure.~\ref{fig:write_latency} compares the average write latency for different number of clusters (K) on the PubMed abstract data set. In this test, to see the impact of K on latency, we invoke insert and delete operations on the system in a 1:1 ratio. Note that the value of K does not affect the lookup request latency because in the lookup, the request does not go through the model or the dynamic address pool. This test shows that by increasing K, latency decreases because all the items within a cluster become more similar (in terms of hamming distance). So, the new items can be written by replacing old ones with a fewer number of cache line writes, which leads to decreasing latency.

\begin{figure} 
         \centering \includegraphics[width=0.9\columnwidth, scale = 0.8]{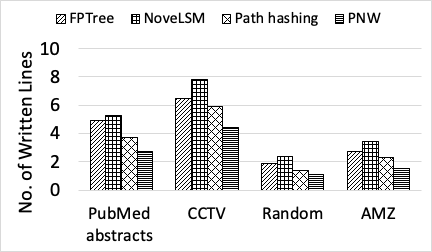}
         \caption{The average number of written cache lines for each request.}
         \label{fig:KV_latency}
\end{figure}

\cut{
Also, this test shows that after some Ks, the write latency does not have a noticeable improvement because we need to write at least a number of cache lines for each request even if a few number of bits are different between the two items.}

In the next test, we compare PNW with recent K/V stores to see its performance in terms of the number of written cache lines. Like the previous test, since only insert and delete requests cause writes to NVMs, we first insert n items into the system and then delete 0.5n items. FP-Tree~\cite{oukid2016fptree} is a hybrid SCM-DRAM persistent B+-Tree method that we implement and compared PNW with. The second persistent K/V store that we compare PNW with is NoveLSM~\cite{kannan2018redesigning}, which is a persistent LSM-based K/V storage system. It is designed to exploit non-volatile memories in an attempt to provide low latency and high throughput to applications. We also implement a hashing scheme that is designed for NVMs called Path hashing~\cite{zuo2017write}. It is worth noting that for this test, we implement PNW as shown in Figure.~\ref{fig:put_delete_DRAM}.

\begin{figure} 
          \centering \includegraphics[width=0.7\columnwidth, angle =0]{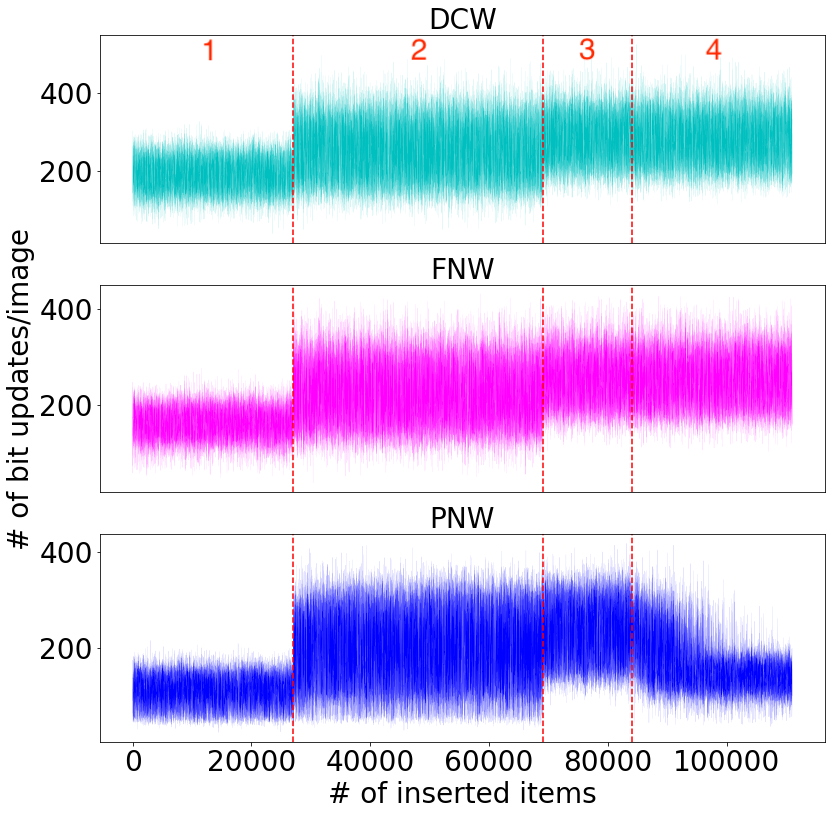}
         \vspace*{-1mm}
         \caption{The performance change by converting the workload from MNIST into Fashion-MNIST over time.}
         \label{fig:fig_10}
\end{figure}

\begin{figure*}
\centering
\begin{subfigure}{.23\textwidth}
\includegraphics[width=\columnwidth]{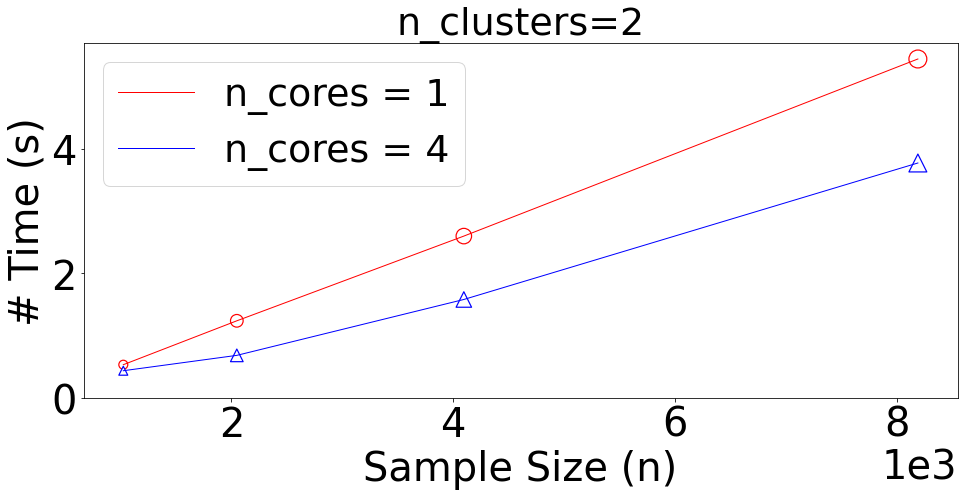}%
\caption{Seq\_2}%
\label{fig:fig_2_seq_core}%
\end{subfigure}\hfill%
\hspace*{-3mm}
\begin{subfigure}{.23\textwidth}
\includegraphics[width=\columnwidth]{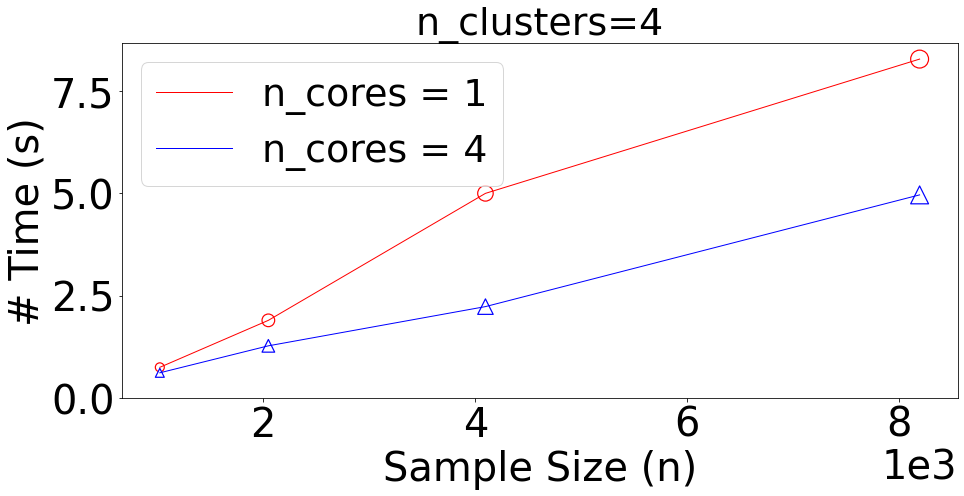}%
\caption{Seq\_4}%
\label{fig:fig_4_seq_core}%
\end{subfigure}\hfill%
\hspace*{-3mm}
\begin{subfigure}{.23\textwidth}
\includegraphics[width=\columnwidth]{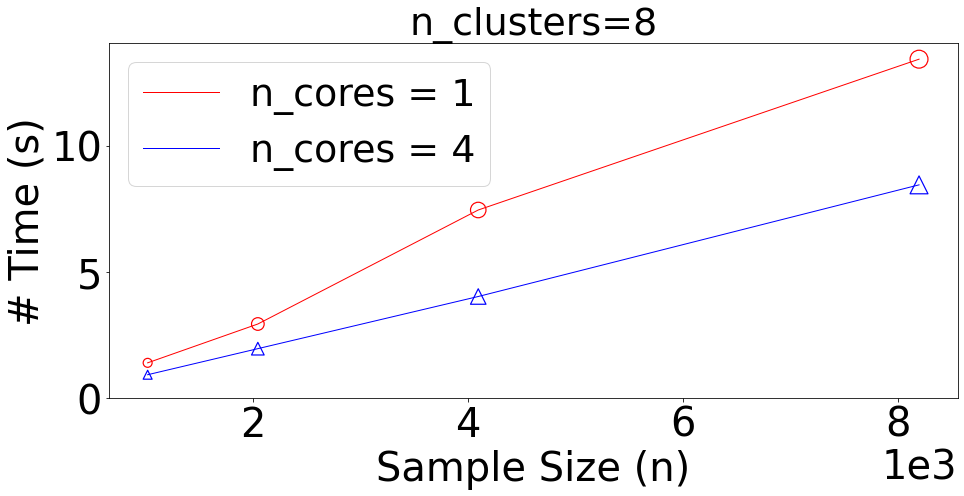}%
\caption{Seq\_8}%
\label{fig:fig_8_seq_core}%
\end{subfigure}\hfill%
\hspace*{-3mm}
\begin{subfigure}{.23\textwidth}
\includegraphics[width=\columnwidth]{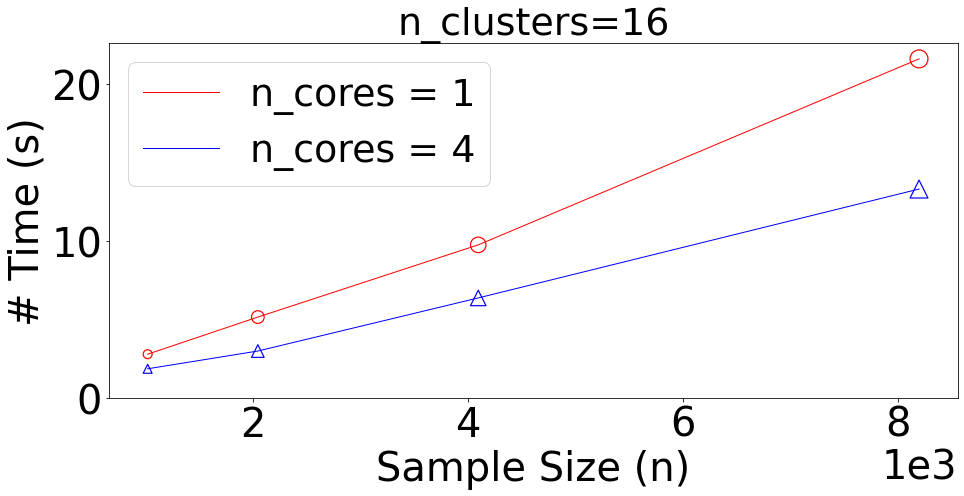}%
\caption{Seq\_16}%
\label{fig:fig_16_seq_core}%
\end{subfigure}\\
\begin{subfigure}{.23\textwidth}
\includegraphics[width=\columnwidth]{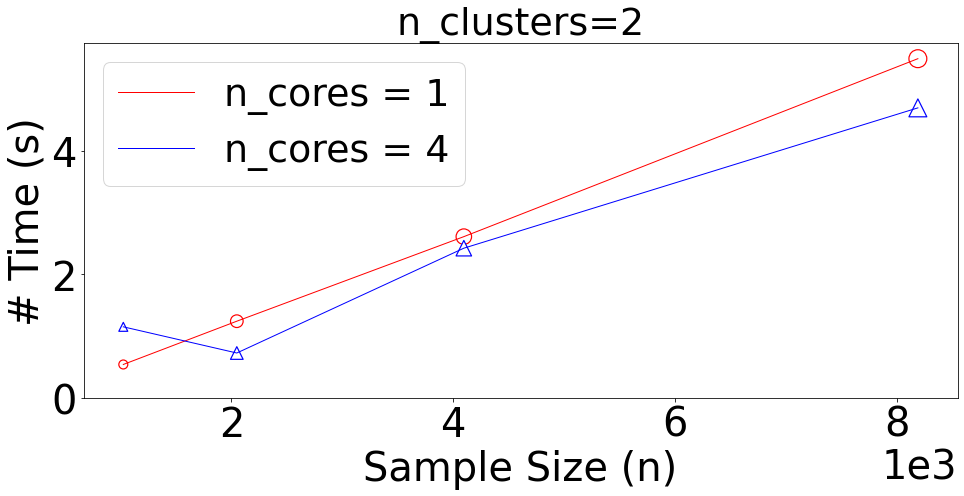}%
\caption{Sher\_2}%
\label{fig:fig_2_sher_core}%
\end{subfigure}\hfill%
\begin{subfigure}{.23\textwidth}
\includegraphics[width=\columnwidth]{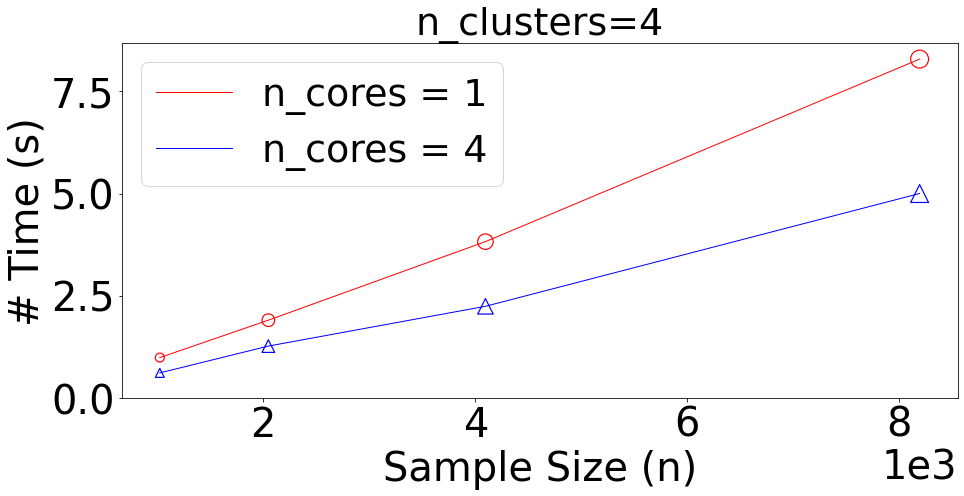}%
\caption{Sher\_4}%
\label{fig:fig_4_sher_core}%
\end{subfigure}\hfill%
\begin{subfigure}{.23\textwidth}
\includegraphics[width=\columnwidth]{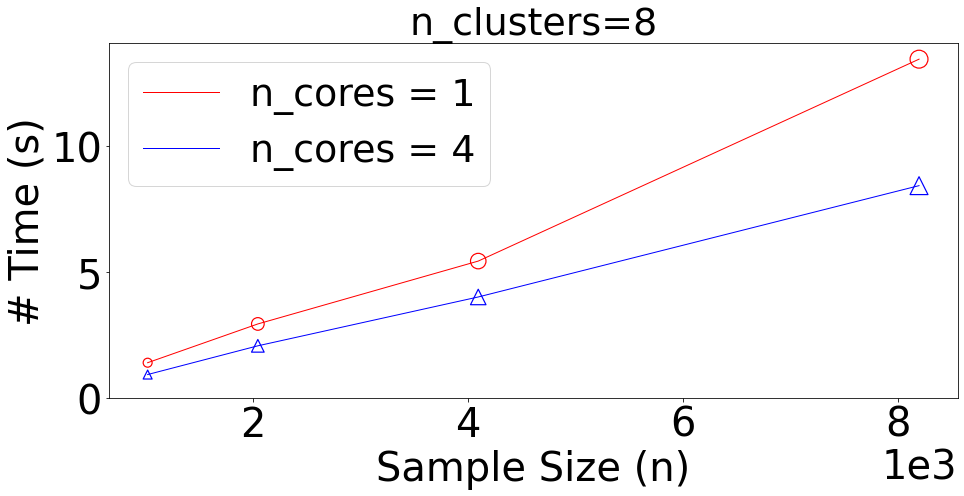}%
\caption{Sher\_8}%
\label{fig:fig_8_sher_core}%
\end{subfigure}\hfill%
\begin{subfigure}{.23\textwidth}
\includegraphics[width=\columnwidth]{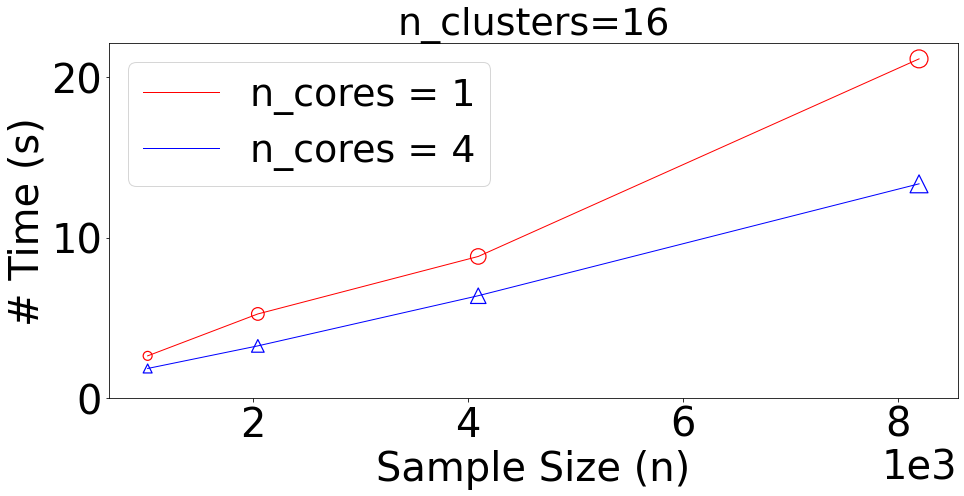}%
\caption{Sher\_16}%
\label{fig:fig_16_sher_core}%
\end{subfigure}
\caption{{\sysname}'s average model training time for different data sets using single core versus multi-core processing.}
\label{fig:fig_cores}
\end{figure*}
Figure.~\ref{fig:KV_latency} shows the average number of written cache lines for each request. The number of written cache lines per request in FPTree and NoveLSM is higher than others because they modify more items to process a request. Although the number of written lines in path hashing is fewer than the others, its written lines are higher than PNW because: 1) It incurs more writes when re-hashing to handle conflicts, and 2) like other methods, it is not ``memory-aware''. PNW has the fewest written cache lines mostly because it can save some cache lines per request because of replacing the old items by similar new items. We also observe that for some data sets the average number of written cache lines is higher for all methods because of the larger item size.

\subsection{Training overhead} \label{Training overhead}
\label{section:Training overhead}
To see how rapidly can our method adapt to changing workloads, we conduct the last experiment to track the behavior of our method while changing the workload. You can see the results in Figure.~\ref{fig:fig_10}. In this test, we use two data sets from the Keras library, i.e., MNIST database of handwritten digits and Fashion-MNIST database of fashion articles, each of which contains 60,000 28x28 gray scale images, along with a test set of 10,000 images. For this test, we did the follows steps:

\begin{itemize}
    \item Phase 1: we stored 28K images from the MNIST data set as the old data. After training the model and creating the dynamic address pool, we started streaming 27K images from the same data set (MNIST) as the new data into the system to overwrite the old data. As we can see in Figure.~\ref{fig:fig_10}, there is no noticeable change in the performance of the system in the first 27K frames. Even at the end of this stage, where the old data is almost completely replaced with the new one, we still do not see any substantial change in the performance.
    \item Phase 2: we send a mixture of items from two different data sets, i.e., Fashion-MNIST and MNIST, at the ratio of 2 to 1. We shuffled 15K of MNIST images with 30K of Fashion-MNIST and then sent them to the system as the new incoming data. As it is obvious in the figure, the performance is affected immediately (the number of updated bits increases) since two-third of the incoming data are entirely different from the previous ones and as a result have a larger hamming distance.
    \item Phase 3: In this phase, we sent 12K images only from the second data set, i.e., Fashion-MNIST. The number of updated bits fluctuated less since the old data contains the items mostly from Fashion-MNIST, and the incoming data is also from the same one too.
    \item Phase 4: In this phase, we continued sending 28K images from the second data set (Fashion-MNIST) with one difference: we re-trained our model on the old data, which contains the images from the Fashion-MNIST data set now. As you can see in the figure, the results got better and fluctuated less.  
\end{itemize}

As a result, we have seen that, depending on the application and the workload, we do not always have to re-train the model rapidly, and we can use the same model for a certain amount of time before it needs to be re-trained. This allows us to do the retraining in the background lazily and update the model periodically.

{\sysname} is designed to enable re-training in the background while the current model is serving requests. However, to set the load factor to its correct value, {\sysname} needs to know when to start re-training the model before the old one becomes inefficient, i.e. the system's performance decreases in terms of the number of bit flips. This is of great importance because we might not want to give all the available resources to the model since the system needs to serve the requests without any problem while the new model is being re-trained. We performed additional experiments to evaluate the costs for re-training a new model using different number of the available cores (Figure.~\ref{fig:fig_cores}). These experiments are performed on the traffic surveillance \cite{bahnsen2018rain} and the Sherbrooke video data sets \cite{jodoin2014urban}. 
\cut{
Furthermore, to run the tests, we used a regular system without any GPU support to have better understanding of {\sysname}'s performance on simple configurations. 
}

In this test (Figure.~\ref{fig:fig_cores}), we calculate the time needed for re-training the model for 2, 4, 8, and 16 clusters. In each case, we did the test on two different modes: 1) running the model on a single core; and 2) running the model on all 4 cores. As we can see from the results, as the value of k and the sample size increases, the model needs more time to be re-trained. For instance, for training a model with k=16 clusters on more than 8000 samples/frames (Figure.~\ref{fig:fig_16_seq_core}), we need almost 20 and 13 seconds if we use one and 4 cores, respectively. This can give us an idea of setting the load factor in a way that we have enough time to finish re-training the new model before the old model becomes inefficient. 
\cut{
Also, in some tests, such as Figure.~\ref{fig:fig_2_sher_core}, the average training time for one core is less than that for multi-core. The reason is that the cost involved with distributing the sub-processes across the different cores is not worth it to distribute the process across them.} So, if we have more than one core available for us in the system to train the model, multi-core processing is worth it when the sample size is big enough.

\subsection{Wear-leveling} \label{wear_leveling}
Aside from decreasing the number of writes, wear-leveling is equally important to extend the lifetime of PCM. The reason is that some blocks of PCM may receive a much higher number of writes than the other blocks, and as a result, wear out sooner \cite{mittal2015survey, xia2015survey}. Therefore, to observe the performance of {\sysname} in terms of the distribution of the maximum number of bit flips and the wear-leveling of PCM, we conduct two more tests. In these tests, we run {\sysname} in two different modes, i.e. for k =5 and k = 30 clusters, on the combination of MNIST and Fashion-MNIST data sets. Like the previous test, we first warm up the data zone with 28K items from the combination of both data sets. Then, we stream 112K writes from the same data sets to the system. During the test, we also perform delete actions to make space for incoming writes. In other words, each word in the data zone is updated 4 times on average.

\begin{figure} %
\centering
\begin{subfigure}{.24\textwidth}
\includegraphics[width=\columnwidth]{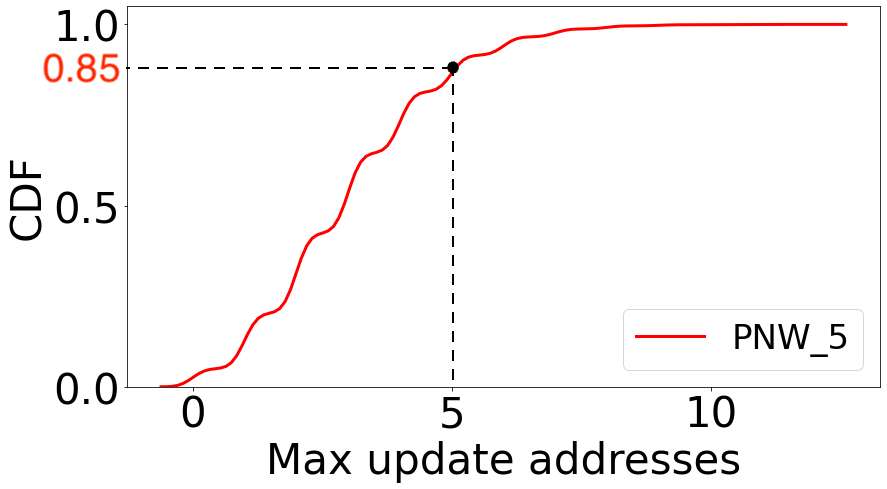}%
\caption{k = 5}%
\label{fig:CDF_5}%
\end{subfigure}
\begin{subfigure}{.24\textwidth}
\includegraphics[width=\columnwidth]{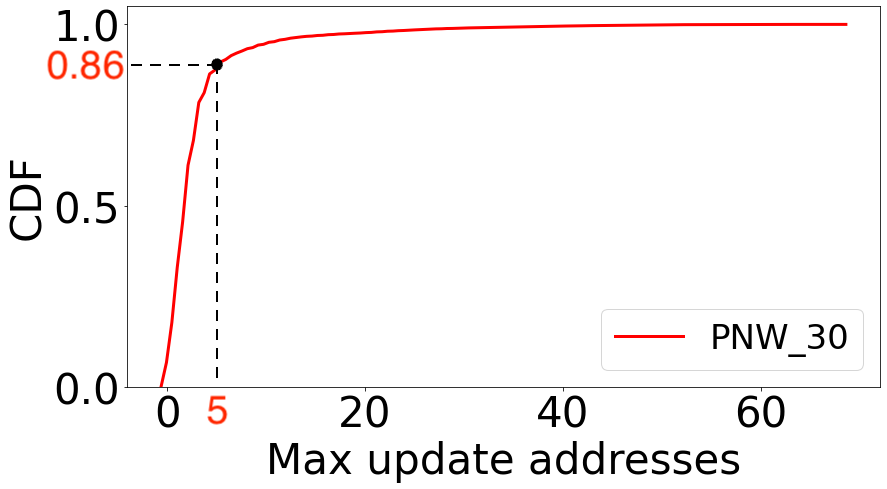}%
\caption{k = 30}%
\label{fig:CDF_30}%
\end{subfigure}\\
\caption{The maximum update addresses as CDFs by applying {\sysname} with a) k=5 and b) k=30 clusters.}
\label{fig:CDF}
\end{figure}

Figure.~\ref{fig:CDF} shows the maximum number of times the addresses in the data zone are written as a cumulative distribution function (CDF). In other words, this figure illustrates the estimation of the likelihood to observe an address in the data zone of PCM that is written less than or equal to a specific number of times. For example, as we can see in Figure.~\ref{fig:CDF_5}, the estimated likelihood to observe an address in the PCM data zone to be written less than or equal to 5 (P (X $\leq$ 5)) is 85\% (Figure.~\ref{fig:CDF_5}) and 86\% (Figure.~\ref{fig:CDF_30}) when we have k = 5 and k = 30 clusters, respectively. We also observe that more than 99\% of the addresses in the data zone experience no more than 10 writes for k = 5 and 15 writes for k = 30. This results show that, regardless of the number of clusters, {\sysname} distributes write activities across the whole PCM chip.

\begin{figure} %
\centering
\begin{subfigure}{.24\textwidth}
\includegraphics[width=\columnwidth]{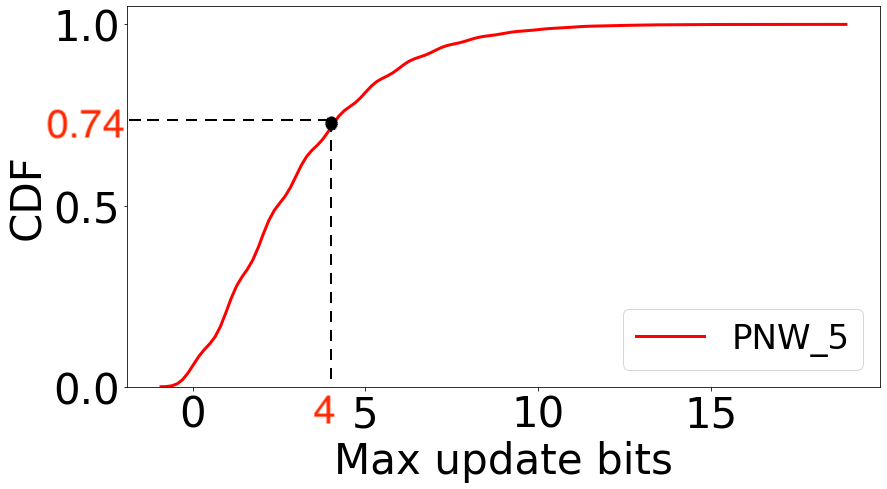}%
\caption{k = 5}%
\label{fig:CDF_bit_5}%
\end{subfigure}\hfill%
\begin{subfigure}{.24\textwidth}
\includegraphics[width=\columnwidth]{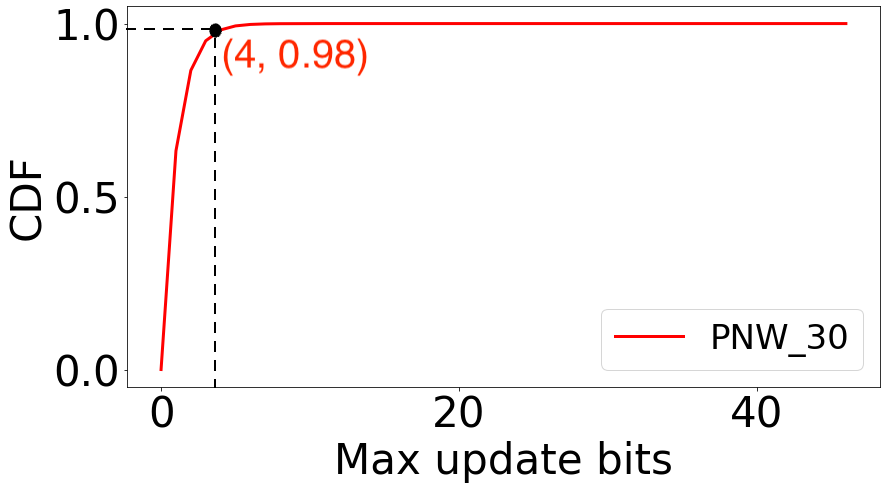}%
\caption{k = 30}%
\label{fig:CDF_bit_30}%
\end{subfigure}\\
\caption{Wear-leveling as CDFs by applying {\sysname} with a) k=5 and b) k=30 clusters.}
\label{fig:CDF_bit}
\end{figure}
Finally, we analyze the wear-leveling of memory bits as CDFs. Figure.~\ref{fig:CDF_bit} illustrates the estimation of the likelihood to observe a memory bit in the data zone of PCM that is written less than or equal to a specific number of times. For instance, we observe that while the estimated likelihood of a memory bit being written less than or equal to 4 times is 74\% for k=5 clusters (Figure.~\ref{fig:CDF_bit_5}), this likelihood rises to 98\% when k=30 (Figure.~\ref{fig:CDF_bit_30}). This important observation shows an interesting fact about {\sysname}: By increasing the number of clusters, bit flips are distributed more evenly across the whole data zone of the PCM chip, and as a result, the lifetime of PCM is extended more. The reason behind this is that when the number of clusters increases, the items within the clusters become more similar to each other. Therefore, regardless of the number of clusters, {\sysname} evenly distributes writes not only in the address level but also in the bit level.

\cut{
At the end of this section, we can conclude that by choosing smaller values for k, the re-training process is triggered less frequently since when there are more addresses available to each cluster in the dynamic address pool and the system reaches to its load factor later. However, smaller k also means that we are sacrificing the wear-leveling in the NVM to get the mentioned advantages. Therefore, it is the design choice of the application to set these hyper-parameters in such a way to get the desired results. 
}

\section{Conclusion}
\cut{
Building storage systems, such as K/V stores, on hybrid memory, creates unprecedented opportunities to utilize fast memory access to achieve enhanced performance compared to those on traditional hard disks or flash-based solid-state drives (SSDs).}

In this paper, we improve the write bandwidth, write energy, write latency, and write endurance of NVMs 
\cut{with machine learning.}
through \emph{Predict and Write} ({\sysname}), a K/V store that uses a clustering-based approach to extend the lifetime of NVMs using machine learning. 
\cut{
To this end, we bring a well-known unsupervised machine learning algorithm, called K-means clustering, into NVMs territory. In our proposed method, we have built the model on DRAM based on the existing old data on PCM. In this way, we try to prevent data items from being moved on PCM by making the model change, which is on DRAM. In other words, by bringing all the extra writes from NVM to DRAM, {\sysname} does not impose any extra writes to NVM. }
We examined the performance of our proposed approach with others in terms of different factors such as the number of writes and the latency for various workloads, on both synthetic and real-world data, with different distributions of data. The results show that our method outperform existing solutions and that the benefit of using a ML model outweigh its overhead. Based on the results, by choosing the right target memory location for a given PUT/UPDATE operation, {\sysname} has succeeded in reducing the number of total bit flips and cache lines over the state of the art.

\section{Acknowledgments}
This research is supported in part by the NSF under grants CCF-1942754 and CNS-1815212.

\bibliographystyle{IEEEtran}
\bibliography{mybibfile}

\end{document}